\documentclass[twocolumn,english,showpacs,preprintnumbers,amsmath,amssymb,floatfix,nofootinbib]{revtex4-1}

\usepackage{xcolor}
\usepackage[T1]{fontenc}
\usepackage[latin9]{inputenc}
\usepackage{color}
\usepackage{array}
\usepackage{amstext}
\usepackage{graphicx}
\usepackage{esint}
\usepackage{rotating}
\usepackage[font=small,labelfont=bf]{caption}
\usepackage{xcolor}
\usepackage[colorlinks = true,
linkcolor = magenta,
urlcolor  = blue,
citecolor = red,
anchorcolor = blue]{hyperref}

\usepackage{ulem}

\makeatletter

\begin{document}

%
%
\title{ NNLO charmed-meson fragmentation functions and their uncertainties in the presence of meson mass corrections } 
%
%

\author{Maral Salajegheh$^{1}$}
\email{Maral.Salajegheh@gmail.com}

\author{S. Mohammad Moosavi Nejad$^{1,4}$}
\email{Mmoosavi@yazd.ac.ir}

\author{Maryam Soleymaninia$^{4}$}
\email{Maryam\_Soleymaninia@ipm.ir}

\author {Hamzeh Khanpour$^{3,4}$}
\email{Hamzeh.Khanpour@cern.ch}

\author {S. Atashbar Tehrani$^{4}$}
\email{Atashbar@ipm.ir}

\affiliation {
$^{1}$Physics Department, Yazd University, P.O.Box 89195-741, Yazd, Iran    \\
$^{3}$Department of Physics, University of Science and Technology of Mazandaran, P.O.Box 48518-78195, Behshahr, Iran     \\
$^{4}$School of Particles and Accelerators, Institute for Research in Fundamental Sciences (IPM), P.O.Box 19395-5531, Tehran, Iran  }

\date{\today}

%
\begin{abstract}

The main aim of this paper is to present new sets of non-perturbative fragmentation functions (FFs) for $D^0$ and $D^+$ mesons at next-to-leading (NLO) and, for the first time, at next-to-next-to-leading order (NNLO) in the $\overline{\mathrm{MS}}$ factorization scheme with five massless quark flavors. This new determination of FFs is based on the QCD fit to the {\tt OPAL} experimental data for hadron production in the electron-positron single-inclusive annihilation (SIA). We discuss in detail the novel aspects of the methodology used in our analysis and the validity of obtained FFs by comparing with previous works in literature which have been carried out up to NLO accuracy. We will also incorporate the effect of charmed meson mass corrections into our QCD analysis and discuss the improvements upon inclusion of these effects. The uncertainties in the extracted FFs as well as in the corresponding observables are estimated using the ``Hessian'' approach. For a typical application, we use our new FFs to make theoretical predictions for the energy distributions of charmed mesons inclusively produced through the decay of unpolarized top quarks, to be measured at the CERN LHC.
As a result of this analysis, suggestions are discussed for possible future studies on the current topic to consider any theory improvements and other available experimental observables.

\end{abstract}


\maketitle

\tableofcontents{}

%
\section{Introduction}\label{sec:introduction}
%

Previous measurements of charmed mesons production at high energy scattering experiments~\cite{Acosta:2003ax} have shown that inclusive heavy quarks production provides an interesting test for the strong interaction dynamics.
Historically, the first results for the charm quark photoproduction cross section $\sigma(\gamma p\to c\bar{c}+X)$ were presented by ZEUS~\cite{Derrick:1995sc} and H1~\cite{Aid:1996hj} Collaborations at $ep$ collider HERA.
With the advent of high statistics data from the CERN Large Hadron Collider (LHC), one can expect that it allows to test the predictions of QCD with much higher precision, however, this is not only important for testing QCD itself; but a better understanding of heavy quarks productions is also vital for searches of new physics (NP) phenomena and beyond standard model (BSM) researches~\cite{CidVidal:2018eel,Azzi:2019yne}.

Over the last few years, several experimental collaborations at   $p \bar{p}$ and $ep$ colliders have presented data on the differential cross section $d^2\sigma/(dydp_T)$ for the inclusive production of $D^0$, $D^+$ and $D_s^+$ mesons~\cite{Acosta:2003ax,Aktas:2004ka,Breitweg:2000qi}, where $y$ and $p_T$ refer to the rapidity and transverse momentum of mesons, respectively. On the theoretical side, fragmentation functions (FFs) describing the transition probability of the initial partons into observed hadrons, are needed as nonperturbative inputs for the calculation of all the relevant cross sections measurements~\cite{Bertone:2018ecm,Bertone:2017tyb,deFlorian:2007ekg,Kniehl:2000fe,Ethier:2017zbq,deFlorian:2007aj}.
In the framework of QCD, FFs encode the long-distance dynamics of the interactions among quarks and gluons which lead to their hadronization in a hard-scattering process. Generally speaking, to obtain theoretical predictions for all observables involving identified hadrons in the final state, FFs have to be convoluted with partonic cross-sections encoding the short-distance dynamics of  hadron production process.

In Ref.~\cite{Kniehl:2006mw}, non-perturbative $D^0$, $D^+$ and $D^+_s$ FFs are determined both at leading order (LO) and next-to-leading order (NLO) in the modified minimal subtraction ($\overline{\mathrm{MS}}$) factorization scheme by fitting the fractional energy spectra of these mesons measured by the {\tt OPAL} Collaboration~\cite{Alexander:1996wy,Ackerstaff:1997ki} in electron-positron annihilation on the Z-boson resonance at the CERN LEP1 collider. Authors in Ref.~\cite{Kniehl:2006mw} have applied the massless scheme or zero-mass variable-flavor-number scheme (ZM-VFNS) where heavy quarks are treated as any other massless parton.
This scheme can be applied to the open production of heavy flavors, such as $D$ and $B$ hadrons, provided the hard energy scale characteristic for the production process is sufficiently larger than the heavy-flavors mass. This is certainly the case for our applications, because $m_D\ll m_Z, m_t$.
In Ref.~\cite{Kneesch:2007ey}, the authors updated their previous results~\cite{Kniehl:2006mw} using newer data on charmed meson production presented by the {\tt ALEPH}~\cite{Barate:1999bg}, {\tt BELLE}~\cite{Seuster:2005tr} and the {\tt CLEO}~\cite{Artuso:2004pj} Collaborations.
Compared to the data from {\tt ALEPH} and {\tt OPAL}, the data from {\tt BELLE} and {\tt CLEO} are located much closer to the thresholds $\sqrt{s}=2m_c$ and $\sqrt{s}=2m_b$ of the transitions $c\to H_c$ and $b\to H_b$, where $H_c$ and $H_b$ stand for generic $c$ or $b$ hadrons, respectively.  However, at the present, the {\tt BELLE} data included in their analyses have been removed due to an unrecoverable error in the measurements.

In the present analysis, using the ZM-VFN scheme we focus on the hadronization of charm- and bottom-quarks into $D^0$ and $D^+$ mesons  and provide the first QCD analysis of $(c, b) \to (D^0, D^+)$ FFs at NNLO accuracy in perturbative QCD.
We perform a QCD fit to the electron-positron single inclusive annihilation (SIA) data measured by {\tt OPAL} Collaboration. It should be mentioned here that, our analysis is restricted to the SIA data only due to the lack of other theoretical partonic cross sections for single inclusive production of partons at NNLO accuracy. 
We shall also go beyond Refs.~\cite{Kniehl:2006mw,Kneesch:2007ey} by performing a full-fledged error estimation, both for the FFs and the resulting normalized total cross sections, employing the Hessian approach.
Furthermore, we include the effects of meson mass on the FFs, which modify the relations between partonic and hadronic variables and reduce the available phase space. Hadron masses are responsible for the low-energy thresholds.

As an application of our analysis, we apply the extracted FFs to make our theoretical predictions for the energy distribution of charmed mesons produced through the top quark decays. In the SM of particle physics, the top quark is the heaviest elementary particle so that its large mass does not allow it to make a constrained state. Therefore, the top quark  decays before hadronization takes place, i.e.  $t \to b W^+ \to H W^+ + {\text {Jets}}$, where $H$ refers to the observed colorless hadron in the final state. Therefore, at the CERN LHC, the study of energy distribution of observed hadrons can be considered as a channel to indirect search for detailed study of top quark properties. The study of this energy spectrum can be even considered as a new window towards searches on new physics. In fact, any considerable deviation of energy distribution of mesons from the  theoretical predictions in the SM theory can be related to the new physics. In this work, using the obtained results for the charmed meson FFs, we make our predictions for the energy distributions of charmed mesons produced through the decay of unpolarized top quarks at NLO and NNLO accuracies.

The outline of this paper is as follows. In Section~\ref{sec:theor}, we shortly describe our theoretical framework to compute charmed meson FFs. In Section~\ref{sec:mass-correction}, we describe how one should  incorporate the effects of meson mass into the formalism. Section~\ref{sec:data-selection} includes a detailed description of experimental data sets used in this analysis. The results of this study are presented in Sec.~\ref{sec:Results}. The extracted $D^0$ and $D^+$ FFs are discussed in this section and comparisons with other results in literature will be also presented. This section does also include comparisons of analyzed SIA data sets with our NLO and NNLO theoretical predictions for the normalized cross sections along with the one-sigma error bands. In Sec.~\ref{sec:energy-spectrum}, we will present an example for the application of our analysis. Sec.~\ref{sec:conclusion} includes the summary and conclusions.

\section{ Theoretical framework } \label{sec:theor}

In the present work using the ZM-VFN scheme, where the heavy quarks are treated as other massless partons, we  mainly  focus on the hadronization of charm- and bottom-quarks into charmed mesons which are of particular relevance in the era of the LHC~\cite{Kniehl:2012ti,Hamon:2018zqs,Adam:2016ich,Maciula:2013oba,Schweda:2014tya}. In particular, we perform a new QCD analysis of OPAL experimental data~\cite{Alexander:1996wy} for $D^+$ and $D^0$ mesons up to NNLO accuracy.
As usual the optimal way to determine the FFs is to fit them to experimental date extracted from the single-inclusive electron-positron annihilation processes via a virtual photon ($\gamma$) or $Z$ boson, i.e.
\begin{eqnarray}\label{eq1}
e^+ e^- \to
(\gamma, Z)
\to
H_c + X \,,	
\end{eqnarray}
where $X$ stands for the residual final state including unobservable jet and $H_c$ denotes the charmed meson; $D^0$ and $D^+$ in our analysis.
In comparison with the hadron collisions, for example $ep$ or $pp$, the SIA process has less contributions from background processes so one does not need to deal with the uncertainties due to the parton distribution functions (PDFs).  Hence, SIA would provide the cleanest process for the determinations of FFs rather than the $ep$ of $pp$ processes.
The differential cross section of hadron production can be calculated theoretically in the frame work of QCD through the factorization theorem~\cite{Collins:1998rz}. According to this theorem,  the differential cross section of $e^+ e^-$ annihilation can be written as a convolution of hard scattering subprocess, i.e. $e^+e^-\to i\bar{i}$, and the non-perturbative meson FFs describing $i/\bar{i}\to D$, i.e.
\begin{eqnarray}
\label{eq2}
\frac{d\sigma}{dx_{D}}
(e^+
e^- \rightarrow D +X)&=&
\nonumber  \\
&& \hspace{-4.0cm}
\sum_{i=g,u/\bar{u},
\cdots,b/\bar{b}}
\int_{x_D}^{1}
\frac{dy}{y}
D_{i}^{D}
(\frac{x_{D}}{y}, \mu_F)
\frac{d\sigma_{i}}{dy}
(y, \mu_{R}, \mu_{F}) \,.
\end{eqnarray}
Denoting the four-momenta of the virtual gauge boson and the $D$ meson by $q$ and $p_D$, respectively, so that $s=q^2$ and $p_D^2=m_D^2$, in (\ref{eq2}) we introduced the scaling variable $x_D = 2 (p_D\cdot q)/q^2$.
In the center-of-mass (c.m.) frame where $q = (\sqrt{s}, \vec{0})$ this variable is simplified to as $x_D = 2 E_D/\sqrt{s}$ where $E_D$ refers to the energy of D-meson. Thus, $x_D$ represents the energy of $D$-meson in units of the beam energy. In Eq.~(\ref{eq2}), $\mu_F$ and $\mu_R$ are the factorization and renormalization scales, respectively.
The scale $\mu_F$ is associated to the divergences due to collinear gluon radiation off a massless primary quark or antiquark.
To avoid these divergences which are proportional to $\ln(s / \mu_F^2)$ we adopt the usual convention $\mu_F = \mu_R = \sqrt{s}$~\cite{Anderle:2017cgl}.
In Eq.~(\ref{eq2}), $y = 2 (p_i\cdot q)/q^2$ where $p_i$ denotes the four-momentum of parton $i$ so by working in the c.m. frame it is simplified as $y = 2 E_i/\sqrt{s}$.
In fact, $y$ refers to the fraction of the beam energy passed on to the parton $i$. In Eq.~(\ref{eq2}), the Wilson coefficients $d\sigma_i/dy$ are the  differential cross sections at the parton level which can be calculated perturbatively.
Nowadays, these coefficients are known up to NNLO approximation in the perturbative QCD theory, see  for example  Refs.~\cite{Rijken:1996vr,Rijken:1996ns,Mitov:2006wy}.
Since, most of experimental data are presented in the form of $1/\sigma_{\tt tot}\times d\sigma/dx_D$, therefore, to be able to compare our theoretical results with experimental data we need to normalize Eq.~(\ref{eq2}) to the total cross section which reads 
\begin{eqnarray}\label{eq3}
\sigma_{\tt tot}
&=&\frac{4\pi\alpha^2(Q)}{Q^2}
\Big(\sum_i^{n_f}
\tilde{e}_i^2(Q)\Big)
\times \bigg(1+
\alpha_s
K_{\tt QCD}^{(1)}+
\nonumber\\
&&\alpha_s^2
K_{\tt QCD}^{(2)}+
\cdots\bigg),
\end{eqnarray}
where $\tilde{e}_i$ is the effective electroweak charge of quark $i$,  $\alpha$ and $\alpha_s$ are the electromagnetic and the strong coupling constants, respectively.
The coefficients $K_{\tt QCD}^{(i)}$ indicate the perturbative QCD corrections to the LO result and are currently computed up to ${\cal O}(\alpha^3_s)$~\cite{Gorishnii:1990vf}.

In the factorization formula (\ref{eq2}), the $D_i^D$-FFs describe the non-perturbative part of the annihilation process (\ref{eq1}) which are  determined phenomenologically.
In our analysis, to perform the best parametrization we adopt the optimal functional form suggested by Bowler~\cite{Bowler:1981sb} for the parametrization of $c$ and $b$ quark FFs.  It is given by 
\begin{eqnarray}\label{eq4}
D_i^D(z, \mu_{0})
=N_{i} \, z^{-(1 + \beta_i^{2})} \,
(1-z)^{\alpha_{i}} \,
e^{-\beta_i^{2}/z} \, ,
\end{eqnarray}
where $i = c,\overline{c}, b, \overline{b}$. This parameterization form includes three free parameters $\alpha_i$, $\beta_i$ and $N_i$ which should be determined phenomenologically  from the QCD fit to the SIA data sets.
In the proposed form, $\mu_0$ is the initial fragmentation scale which is taken  to be  $\mu_0^2 = 18.5$ GeV$^2$ for heavy flavors $b$ and $c$. The advantage of choosing this amount for the initial scale $\mu_0$, which is a little larger than $m_b^2$, is due to the fact that the time-like matching conditions are currently known up to NLO accuracy and with this input scale the heavy quark thresholds should not be crossed in the QCD evolution.
The FFs of gluon and light flavors $(u, d, s)$ are taken to be zero at the initial scale and are generated at higher scales through the Dokshitzer-Gribov-Lipatov-Alteralli-Parisi (DGLAP) evolution equations~\cite{DGLAP}. For the value of strong coupling constant, we adopt $\alpha_s^{(n_f = 5)}(M_\texttt{Z}) = 0.118$ at the Z boson mass scale~\cite{Tanabashi:2018oca}.  \\
In this analysis, we employ the publicly available {\tt APFEL} package~\cite{Bertone:2013vaa} in order to the evolution of $zD^{D^0}(z, \mu^2)$ and $zD^{D^+}(z, \mu^2)$ FFs as well as for the calculation of the SIA cross sections up to NNLO accuracy.
To determine the free parameters, we applied the CERN program {\tt MINUIT}~\cite{James:1975dr} to minimize the $\chi^2_{\tt global}$ which is defined in our previous work in detail~\cite{Soleymaninia:2019sjo,Soleymaninia:2018uiv}. This quantity includes the overall normalization errors of the D-meson production data sets. The uncertainties are also estimated using the standard Hessian error propagation, as outlined in Refs.~\cite{Stump:2001gu,Martin:2009iq}, with $\Delta \chi^2=1$, which corresponds to a $68\%$ confidence level (CL) in the ideal Gaussian statistics. For more detail, see Ref.~\cite{Soleymaninia:2013cxa}.

\section{Mass corrections } \label{sec:mass-correction}

The factorization theorem described in Eq.~(\ref{eq1}) is routinely used in the literature for the inclusive production of single hadrons.
In proving this formula it is assumed that quarks and hadrons are massless so the non-zero values of the $c$- and $b$-quark masses only enter through the initial conditions of the FFs, and the mass of the heavy baryon sets the lower bounds on the scaling variable $x_H$, i.e. $x_H^{\tt min} = 2m_H/\sqrt{s}$.
In Ref.~\cite{Nejad:2015fdh}, we have proved how to enter the hadron mass effects into the inclusive hadron production in $e^+e^-$ reaction.
To incorporate the hadron mass effects we used a specific choice of scaling variables by working in the light-cone (L.C) coordinates.
Ignoring the detail of calculations, as a generalization of the massless case, the differential cross section in the presence of  hadron mass $m_H$ reads
\begin{eqnarray}
\frac{d\sigma}{dx_H}(x_H, s)
=\frac{1}{1-\frac{m_H^2}
{s\eta^2(x_H)}}
\frac{d\sigma}{d\eta}
(\eta(x_H), s),
\end{eqnarray}
where $\eta = x_H/2 \times (1 - 4m_H^2/(sx_H^2))$ and
\begin{eqnarray}
\frac{d\sigma}{d\eta}(\eta, s)=
\sum_{i}{\int_{\eta}^1{\frac{dy}{y}
\frac{d\hat{\sigma}}
{dy} D_i^H(\frac{\eta}{y}, \mu_F)}}.
\end{eqnarray}
The above formula is a fundamental relation for the factorization theorem extended in the presence of hadron mass and  would be more effective and applicable
when the data are presented in lower energy scales. Among all well-known collaborations, the only collaboration who studied the effects of hadron  mass into their calculations is the  {\tt AKK} collaboration~\cite{Albino:2008fy,Albino:2005me}. Although, due to the data applied in their work, they have identified
the scaling variable $x_p = 2 |\vec{p}|/\sqrt{s}$, where $|\vec{p}|$ stands for the three-momentum of produced hadron, while in our analysis
we use the energy scaling variable as $x_H=2E_H/\sqrt{s}$.  Therefore, the differential cross section or, equivalently, the extended factorization formula including the hadron mass effects computed in our work is different from the one ($d\sigma/dx_p$) presented in Refs.~\cite{Albino:2008fy,Albino:2005me}.

It should be noted that the effect of hadron mass is to increases the cross section $d\sigma/dx_H$ at small $x_H$ so this treatment acts inverse for large $x_H$. We incorporated the mass corrections into the  publicly available {\tt APFEL} package~\cite{Bertone:2013vaa}. Our results show that the inclusion of hadron mass leads to a change  in the value of $\chi^2 /{\tt (d.o.f)}$ up to about $5\%$. Although, due to the scale of energy given for the data used in our work, i.e. $\sqrt{s} = 91.2$~GeV, this improvement  is small but having the data at lower energies it is expected to have more effect on the  $\chi^2 /{\tt (d.o.f)}$.

\section{ Experimental data selection } \label{sec:data-selection}

After introducing the theoretical framework of our analysis in the previous sections, in the following  we discuss the experimental data which are used in this analysis.
The first experimental information corresponding to the D meson production through $e^- e^+$ annihilation comes from the measurements performed by the {\tt OPAL} Collaboration \cite{Alexander:1996wy}. The {\tt OPAL} results have been reported for $D^0$ and $D^+$ production at the Z-boson resonance. In this process, two mechanisms contribute with similar rate;  $Z \rightarrow c\bar{c}$ decay followed by $c/\bar{c}\to H_c$ fragmentation and $Z \rightarrow b \bar{b}$ decay followed by $b/\bar{b}\to H_b$ fragmentation and weak decay of the bottom-flavored hadron $H_b$ into the charmed-meson via $H_b\to H_c+X$. Note that, the energy spectrum of $H_c$ hadrons originating from decays of $H_b$ hadrons is much softer than that due to primary charm production. For separating charmed hadron production through $Z \rightarrow c \bar{c}$ decay from $Z \rightarrow b \bar{b}$ decay, {\tt OPAL} used the apparent decay length distributions and energy spectra of the charmed hadrons.
The  {\tt OPAL} Collaboration has presented $x_D$-distributions for $D^0$ and $D^+$ samples and for their $z\to b\bar{b}$ subsamples (b-tagged events).
They are displayed in the form $1/N_{\tt had}\times dN/dx_D$ where $N$ is the number of charmed meson candidates reconstructed through appropriate decay chains.  To convert these data into the desired cross section  $1/\sigma_{\tt tot}\times d\sigma/dx_D$, one should divide them by the convenient branching fractions of decays used in Refs.~\cite{Alexander:1996wy,Ackerstaff:1997ki} for the reconstruction of the various charmed mesons, i.e.
\begin{align}\label{eq5}
Br (D^{0} \rightarrow
K^{-} \pi^{+})         &=
(3.84 \pm 0.13)\%, \nonumber\\
Br (D^{+} \rightarrow
K^{-} \pi^{+} \pi^{-}) &=
(9.1 \pm 0.6)\%.
\end{align}
These data are displayed in Figs.~\ref{fig:D0-Theory} (for the $D^0$ meson) and \ref{fig:Dp-Theory} (for the $D^+$ meson) and are listed in Tables.~\ref{tab1} and \ref{tab2} along with the corresponding values of $\chi^{2}$. The number of data points and the data properties as well as  the center-of-mass energy ($\sqrt{s}$) are presented in these tables. We have also presented the total $\chi^{2}$ divided by the number of degrees of freedom $\chi^2/{\tt d.o.f}$ for each analysis. These values show the quality of our well-satisfying fit. 
\\
Another important measurement on the charmed mesons production  has been done by {\tt  CLEO} Collaboration~\cite{Artuso:2004pj} in the  $e^-e^+$ annihilation. Excluding the decay products of $B$ mesons, their results are presented for the scaled momentum spectra $d\sigma/dx_{p}$ at $\sqrt{s} = 10.5$~GeV  where $x_{p} = \vert \overrightarrow{p} \vert/ \vert \overrightarrow{p} \vert_{\tt max}$. As was mentioned, the OPAL Collaboration~\cite{Alexander:1996wy,Ackerstaff:1997ki} has presented the cross section as a function of the scaled energy $x_D$. The relation between these two variables reads \cite{Kneesch:2007ey}
\begin{eqnarray}\label{eq6}
x_{p} =
\sqrt{\frac{x_D^{2} -
\rho_{H}}{
1-\rho_H}},
\end{eqnarray}
where $\rho_H = 4 m_H^{2}/s$. Since the variable $x_D$ ranges as $\sqrt {\rho_{H}} <x_D < 1$, consequently $x_{p}$ takes values from 0 to 1. For differential cross section the conversion formula is
\begin{eqnarray}\label{eq7}
\frac{d\sigma(x_p)}{dx_p}
= (1-\rho_{H})
\frac{x_{p}}{x_D}
\frac{d \sigma(x_D)}{dx_D}.
\end{eqnarray}
Although the {\tt CLEO} data can provide useful information on the $D^0$ and $D^+$ FFs, but there is a problem in using them in the analysis. In fact, as was mentioned, we have chosen the ZM-VFN scheme in our analysis which is reliable only in the region of large transverse momenta or equivalently in high energy scales.  On the other hand, the data from {\tt CLEO} are located much closer to the thresholds $\sqrt{s} = 2 m_{c}$ and $\sqrt{s} = 2 m_{b}$ of the transitions $c \to H_{c}$ and $b \to H_{b}$, than those from {\tt OPAL}. Hence, including the {\tt CLEO} data sets in the analysis might be a reason  for tension. To check this point, we converted the {\tt CLEO} data to the desired form  $d\sigma/dx_D$ using Eqs.~(\ref{eq6}) and (\ref{eq7}) and included them into our analysis. We observed that inclusion of these data increased the value of $\chi^{2}$, significantly. Accordingly, we restricted ourselves to the {\tt OPAL} data sets for extracting the charmed mesons FFs and neglected the {\tt CLEO} data sets. We remind that our choice of the massless scheme was due to the lack of massive partonic cross sections at NNLO, currently.

\begin{table}[t!]
	\caption{ The individual $\chi ^2$ values for inclusive and $b$-tagged cross sections
		obtained at NLO and NNLO. The total $\chi ^2$ and $\chi ^2/d.o.f$ fit for $D^0$ are also shown. }
	\begin{tabular}{lcccccc} \hline \hline
		Collaboration & data & $\sqrt{s}$  GeV &  data   & $\chi^2$ ({\tt NLO}) & $\chi^2$ ({\tt NNLO})  \\
		& properties &          &  points &        \\ \hline
		OPAL   & Inclusive  & 91.2 & 13 & 8.36 & 7.08  \\
		& $b$-tagged  & 91.2 &  13 & 14.62 & 14  \\ \hline
		{\bf TOTAL:} & & & 26 &  22.98   & 21.08    \\
		($\chi^{2}$/{ d.o.f}) & & & &1.149 & 1.05         \\\hline\hline
	\end{tabular} \label{tab1}
\end{table}
\begin{table}[h!]
	\caption{ As in Table \ref{tab1}, but for $D^+$. }
	\begin{tabular}{lcccccc} \hline \hline
		Collaboration & data & $\sqrt{s}$  GeV&  data   & $\chi^2$ ({\tt NLO}) & $\chi^2$ ({\tt NNLO}) \\
		& properties &          &  points &        \\ \hline
		OPAL   & Inclusive  & 91.2 & 13 & 7.24 & 6.2  \\
		& $b$-tagged  & 91.2 &  13 & 8.51 & 8.32  \\ \hline
		{\bf TOTAL:} & & & 26 &  15.75   & 14.52    \\
		($\chi^{2}$/{ d.o.f}) & & & & 0.75 & 0.69          \\ \hline\hline
	\end{tabular} \label{tab2}
\end{table}
In next section, we shall present the main findings of our study and discuss on the results. We shall compare our $zD^{D^0}(z, \mu^2)$ and $zD^{D^+}(z, \mu^2)$ FFs with other results in the literature. A detailed comparison of our NLO and NNLO theoretical predictions for the SIA cross sections with the analyzed data sets will be also presented.

%
%
\section{ Results and discussion }\label{sec:Results}

After introducing all ingredients needed for analysis,  we are now in a situation to perform a fit with experimental data sets for each hadron species ($D^0$ and $D^+$). In this section, at first, the numerical results obtained  for the $(c, b)\to D^0/D^+$ FFs through the perturbative QCD analysis are reported at the initial scale $\mu_0$. Next, as an example, we present our results for the FFs  along with their uncertainty bands for the gluon, charm and bottom quarks up to NLO and NNLO corrections at $\mu^2 = 100$ GeV$^2$. Our results for the FFs are compared to the {\tt KKKS08} FFs \cite{Kneesch:2007ey} for D mesons. In addition, our theoretical predictions as well as the error bands for total and $b$-tagged cross sections are compared with the  analyzed SIA experimental data sets. 
\begin{table}[h!]
	\caption{ The optimal values for the input parameters of the $D^0$-FF (\ref{eq4}) at the initial scale $\mu_0^2 = 18.5$~GeV$^{2}$ determined by  QCD analysis of the experimental data listed in Table~\ref{tab1}. }
	\label{tab3}
	\begin{ruledtabular}
		\begin{tabular}{lcll}
			Parameter & & Best values &   \\     & {\tt NLO}  & &  {\tt NNLO} \\ \hline
			$N_c$						& $284.513$ & & $261.214$ \\
			$\alpha_c$ 			    & $1.341$ & & $1.402$ \\
			$\beta_c$ 					& $1.981$ & & $1.953$ \\
			$N_b$						& $13.127$ & & $12.701$  \\
			$\alpha_b$ 				& $3.944$  & & $4.014$ \\
			$\beta_b$ 					& $0.904$  & & $0.891$
		\end{tabular}
	\end{ruledtabular} \label{D0pars}
\end{table}
\begin{table}[h!]
	\caption{As in Table \ref{tab2}, but for $D^+$.}
	\label{tab4}
	\begin{ruledtabular}
		\begin{tabular}{lcll}
			Parameter & & Best values &   \\    & {\tt NLO} & & {\tt NNLO}  \\  \hline
			$N_c$						& $49.817$ & & $44.357$ \\
			$\alpha_c$ 				& $1.20$ &  & $1.253$  \\
			$\beta_c$ 					& $1.841$ & & $1.806$  \\
			$N_b$					    & $11.664$ & & $10.653$  \\
			$\alpha_b$ 			    & $4.308$  & & $4.343$  \\
			$\beta_b$ 					& $1.095$  & & $1.073$
		\end{tabular}
	\end{ruledtabular}\label{D+pars}
\end{table}

\subsection{  Our FF sets in comparison with {\tt KKKS08}  sets} \label{sec:ResultsFFs}

\begin{figure*}[htb]
	\vspace{-0.50cm}
	\resizebox{0.480\textwidth}{!}{\includegraphics{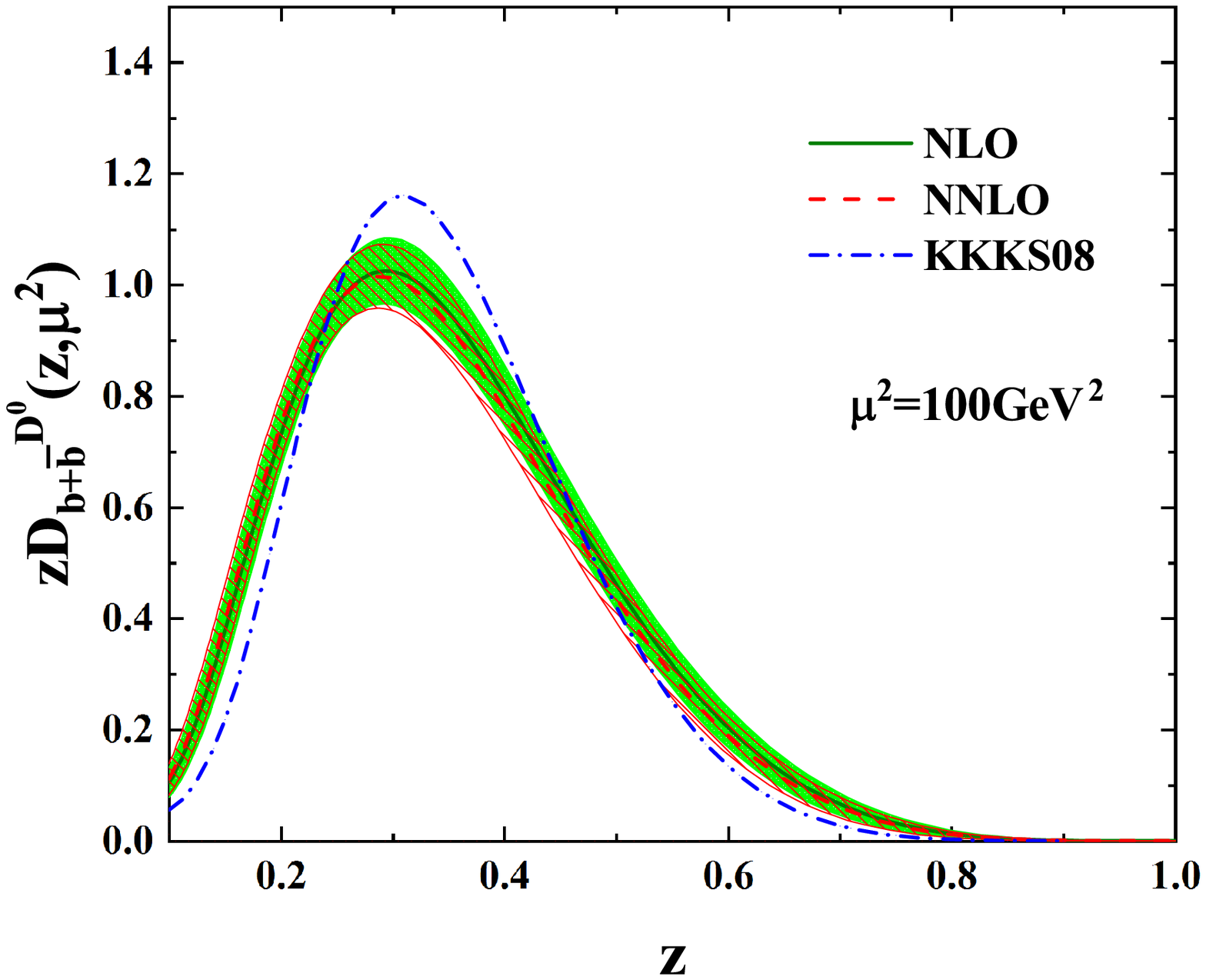}}
	\resizebox{0.480\textwidth}{!}{\includegraphics{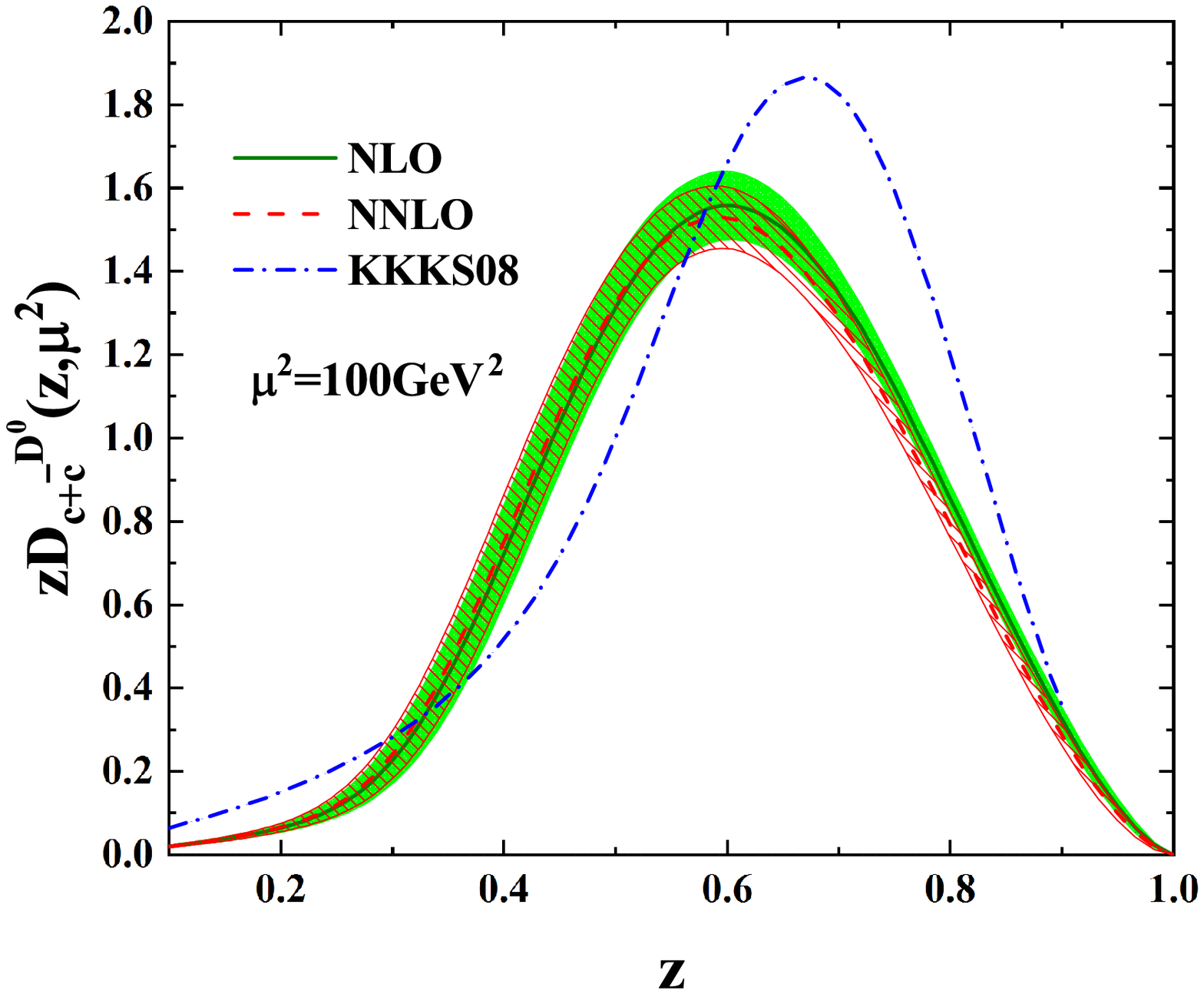}}
	\resizebox{0.480\textwidth}{!}{\includegraphics{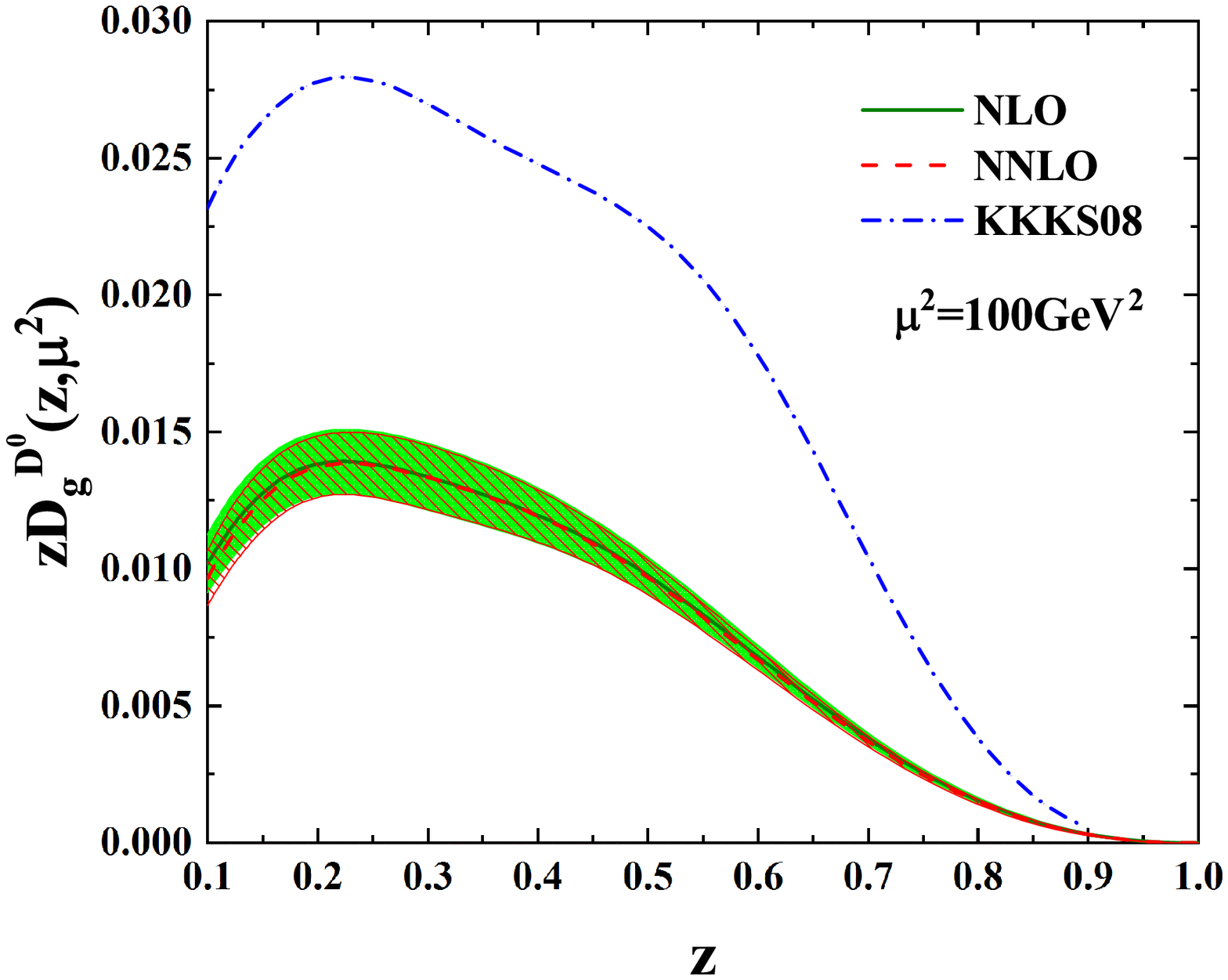}}
	\begin{center}
		\caption{{\small {The $b+\bar{b}$, $c+\bar{c}$ and gluon FFs obtained  at NLO (solid lines) and NNLO (dashed lines) QCD analyses of $D^0$ meson productions. The results of {\tt KKKS08} analysis \cite{Kneesch:2007ey} (dot-dashed lines) are also shown for comparison. The error bands (shaded bands) correspond to the choice of tolerance $\Delta \chi^2 = 1$ obtained by using the Hessian approach.} \label{fig:D0}}}
	\end{center}
\end{figure*}

\begin{figure*}[htb]
	\vspace{0.50cm}
	\resizebox{0.480\textwidth}{!}{\includegraphics{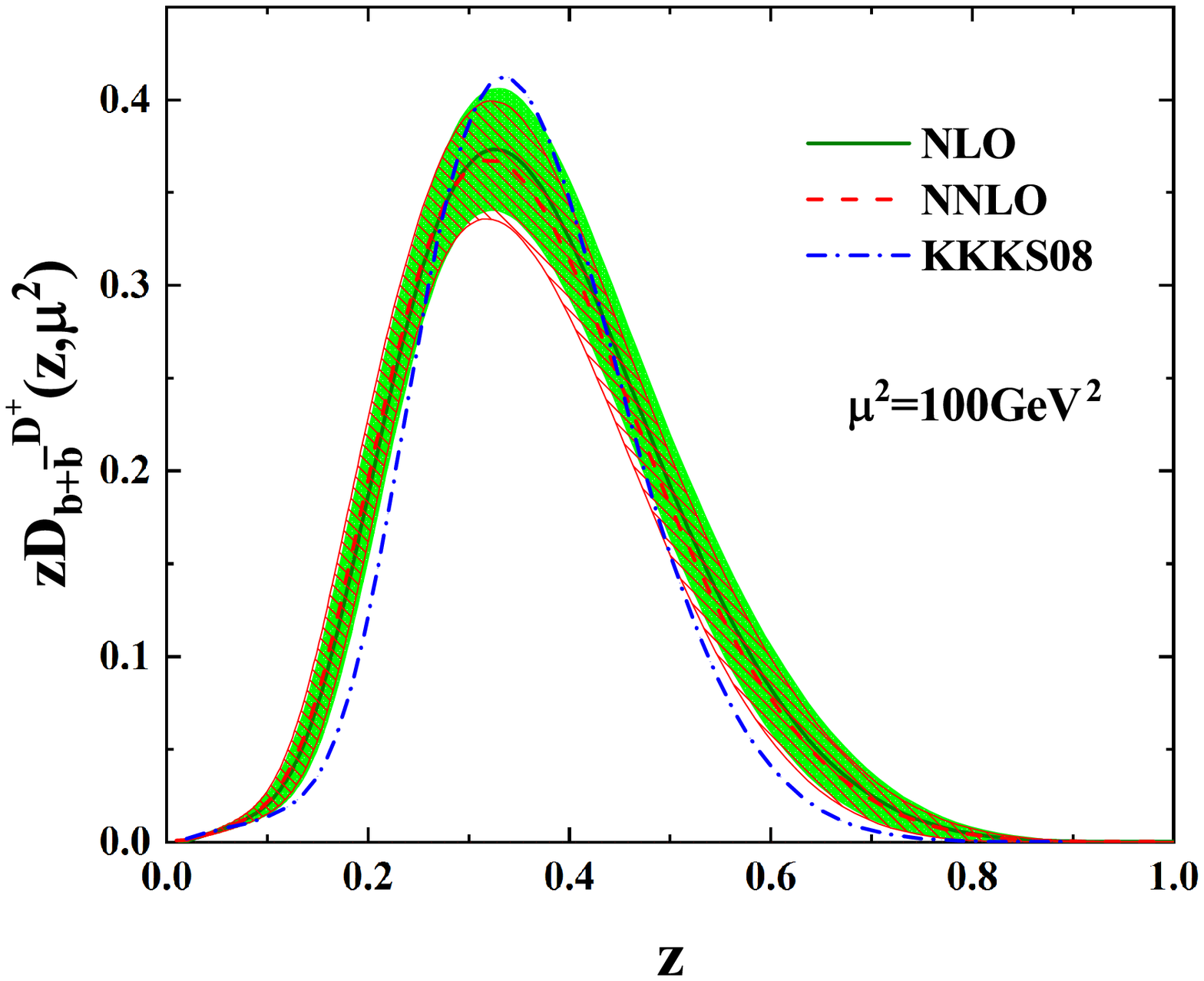}}
	\resizebox{0.480\textwidth}{!}{\includegraphics{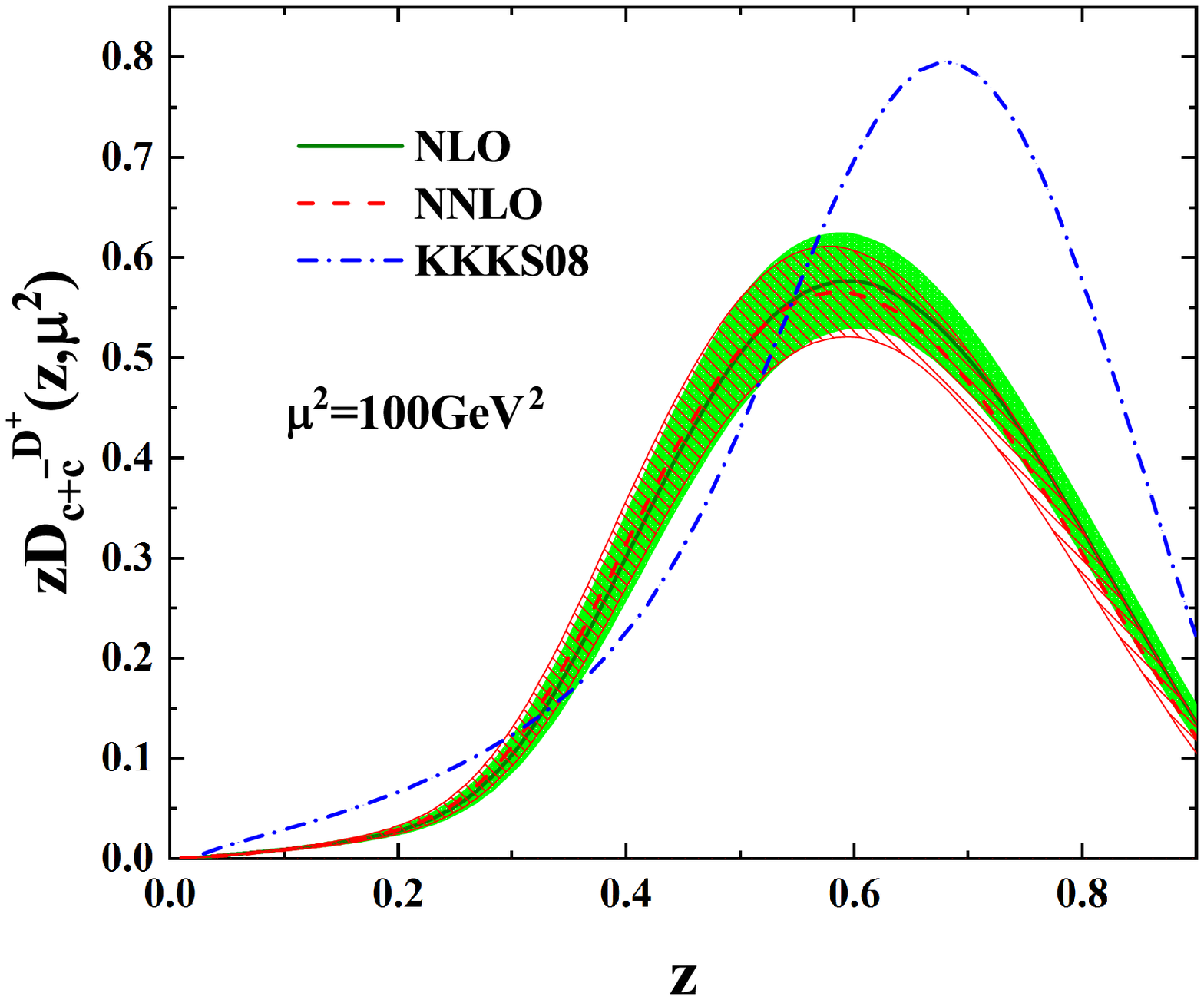}}
	\resizebox{0.480\textwidth}{!}{\includegraphics{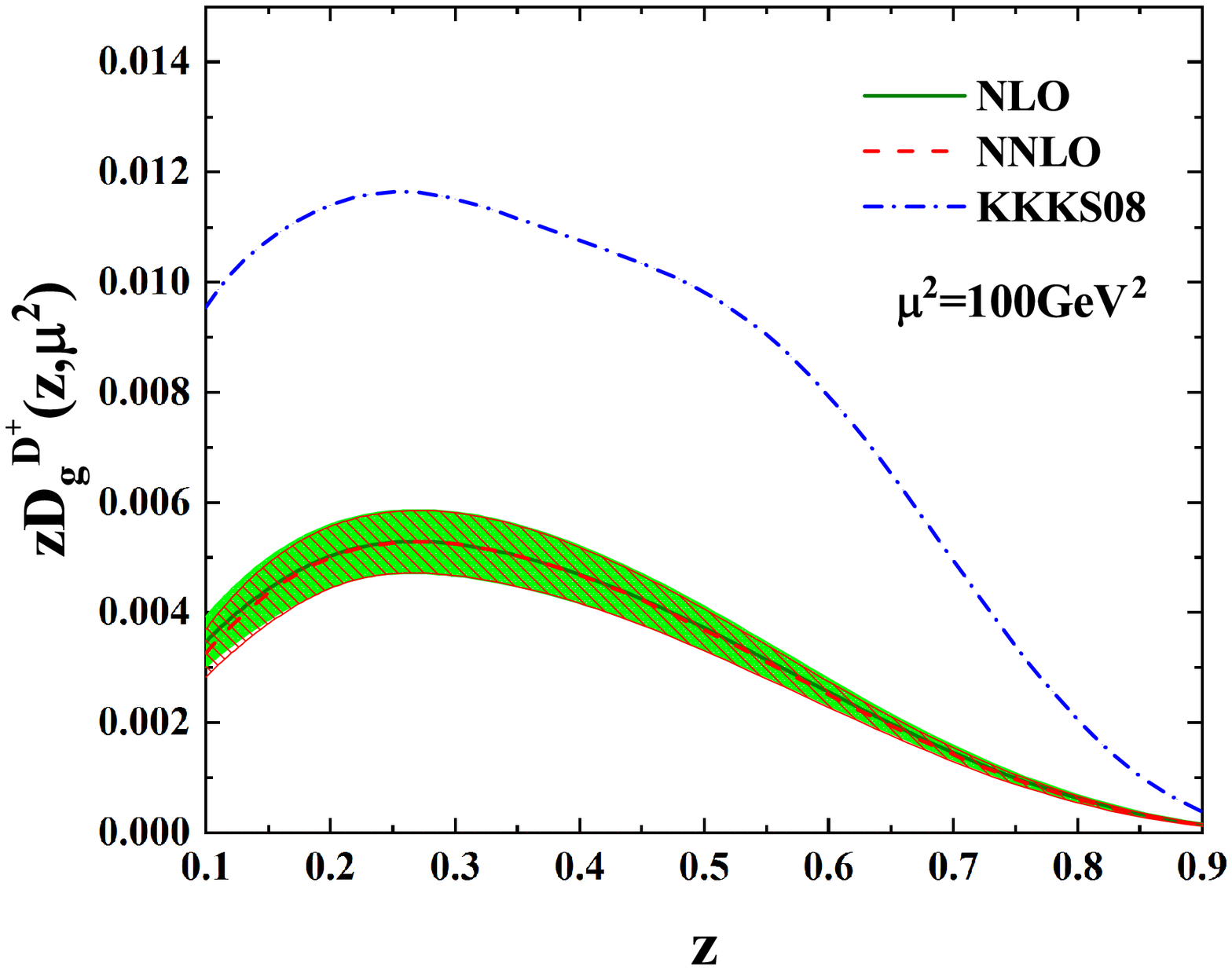}}
	\begin{center}
		\caption{{\small {As in Fig.~\ref{fig:D0}, but for the $D^+$} meson.} \label{fig:DP}}
	\end{center}
\end{figure*}

\begin{figure*}[htb]
	\vspace{0.50cm}
	\resizebox{0.480\textwidth}{!}{\includegraphics{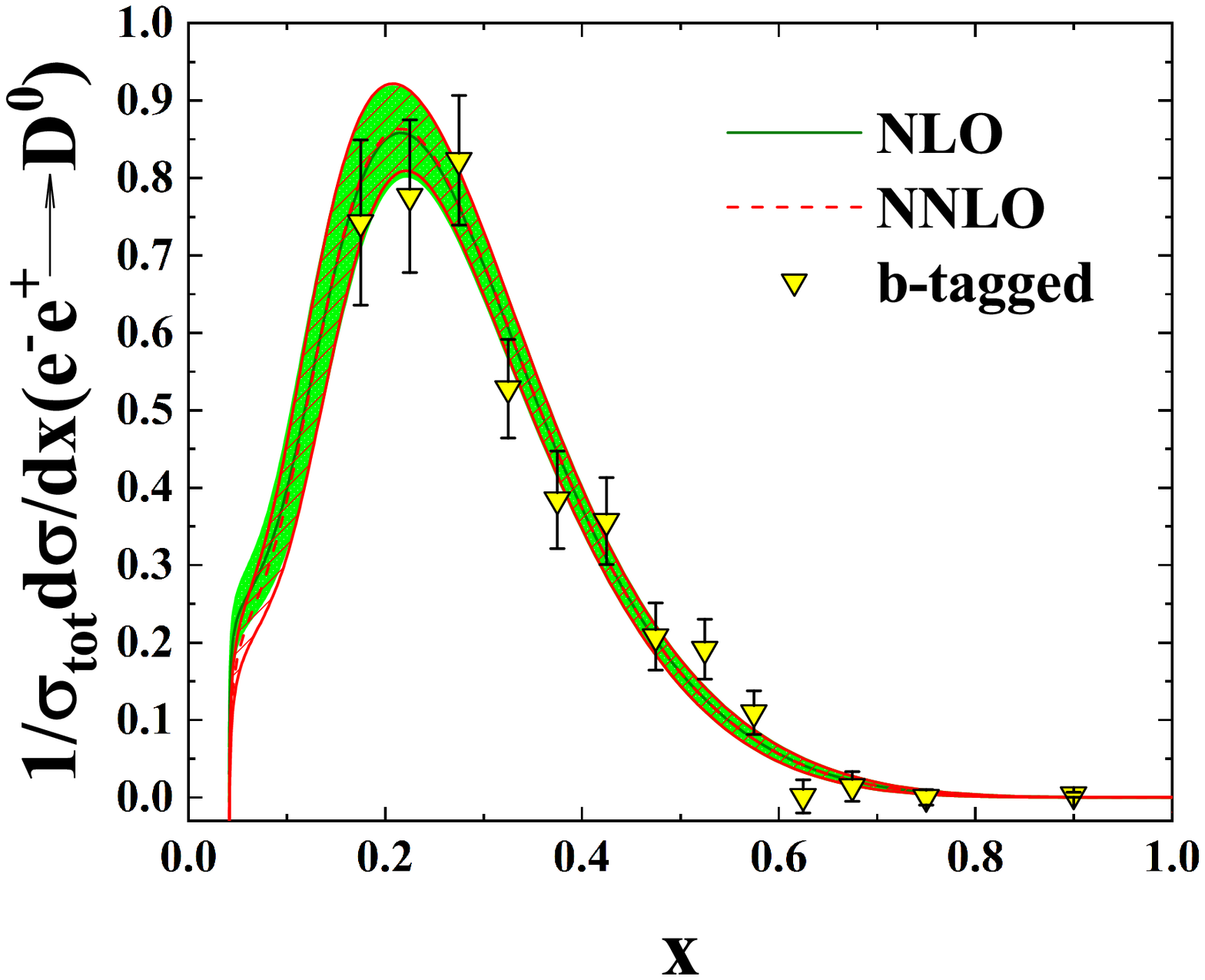}}
	\resizebox{0.480\textwidth}{!}{\includegraphics{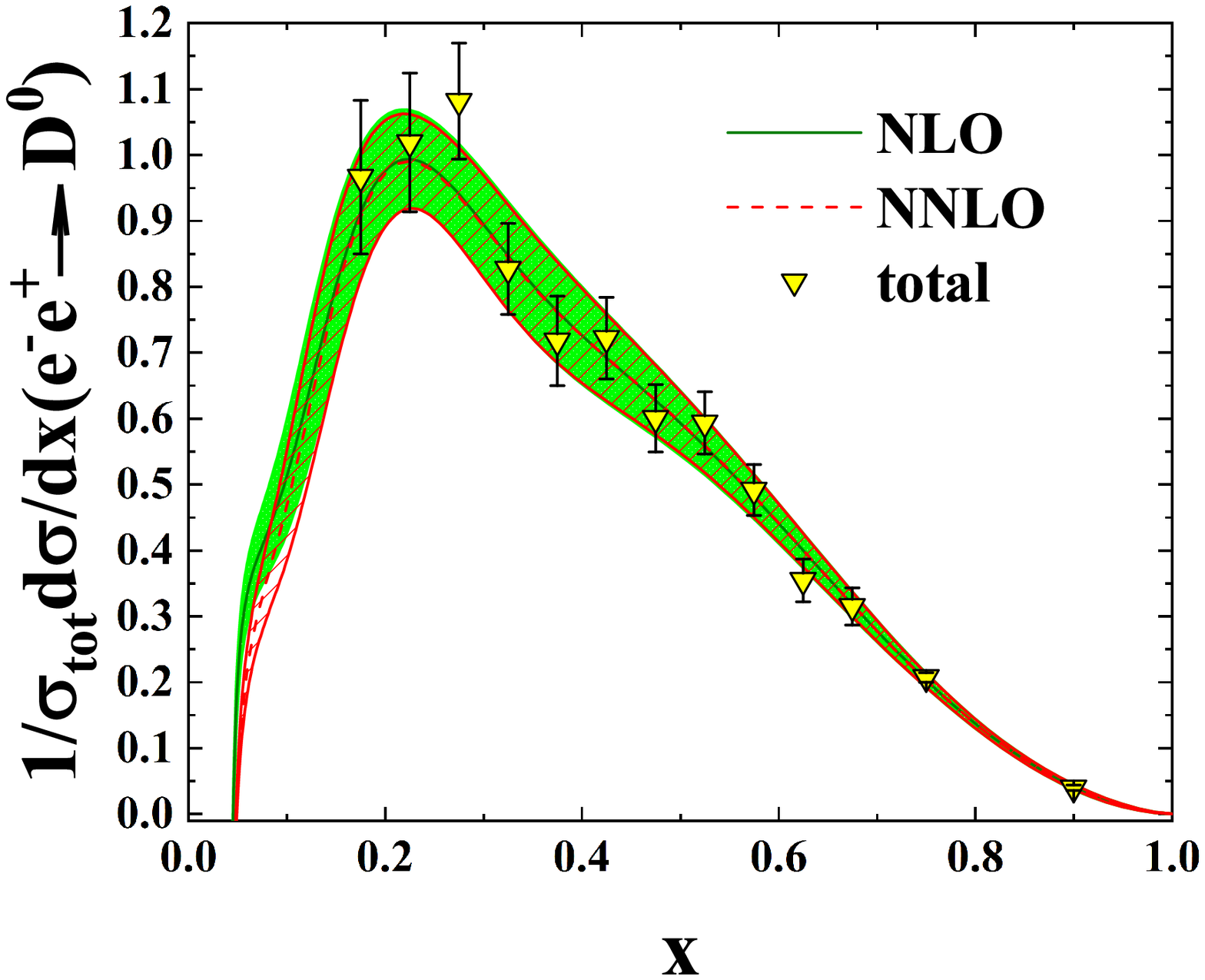}}
	\vspace{-5 cm}
	\begin{center}
		\caption{{\small Our NLO and NNLO theoretical predictions are compared with the normalized inclusive total (left) and $b$-tagged (right) data sets for $D^0$  meson production from  {\tt OPAL} experiment.  } \label{fig:D0-Theory}}
	\end{center}
\end{figure*}
\begin{figure*}[htb]
	\vspace{0.50cm}
	\resizebox{0.470\textwidth}{!}{\includegraphics{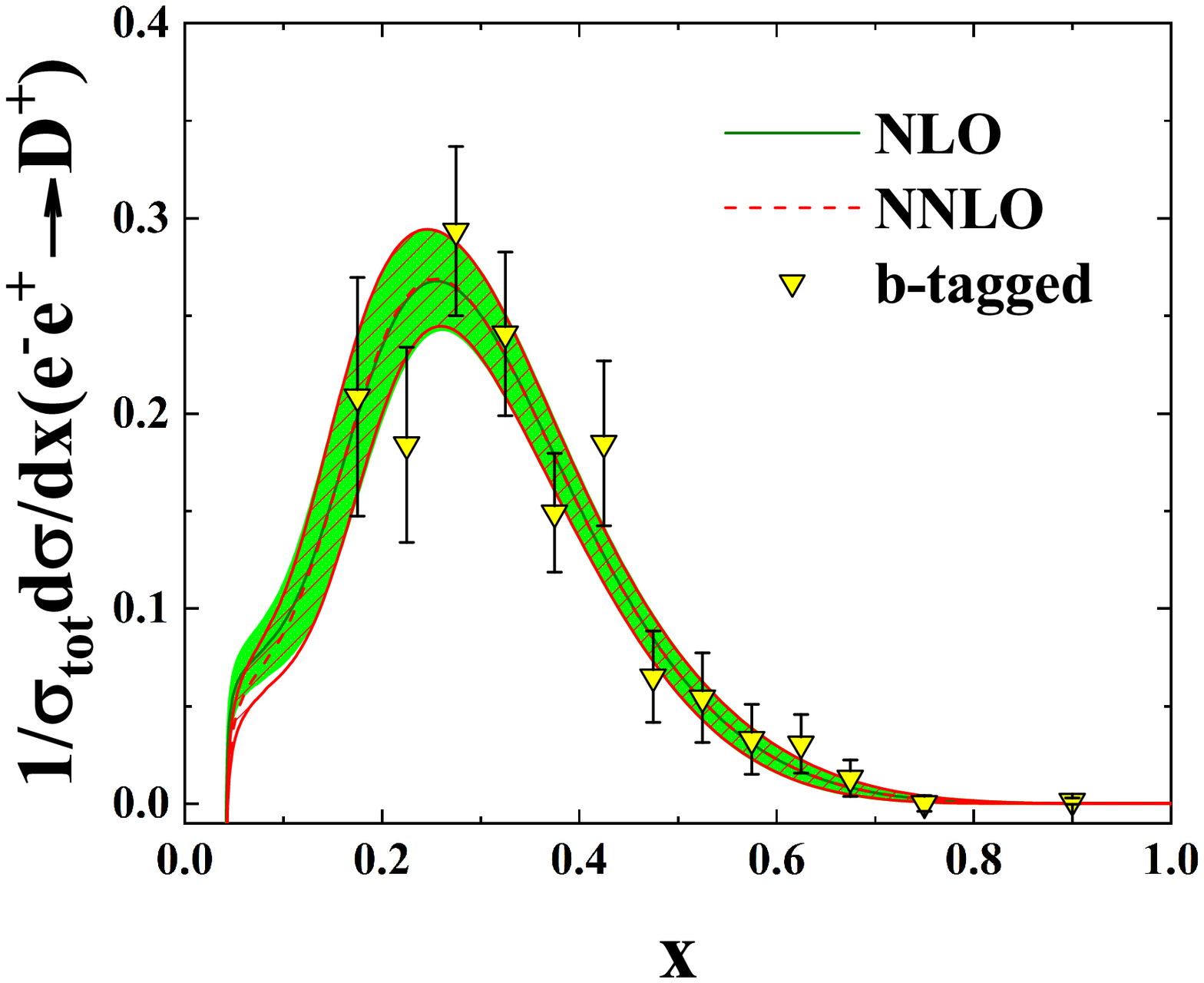}}
	\resizebox{0.470\textwidth}{!}{\includegraphics{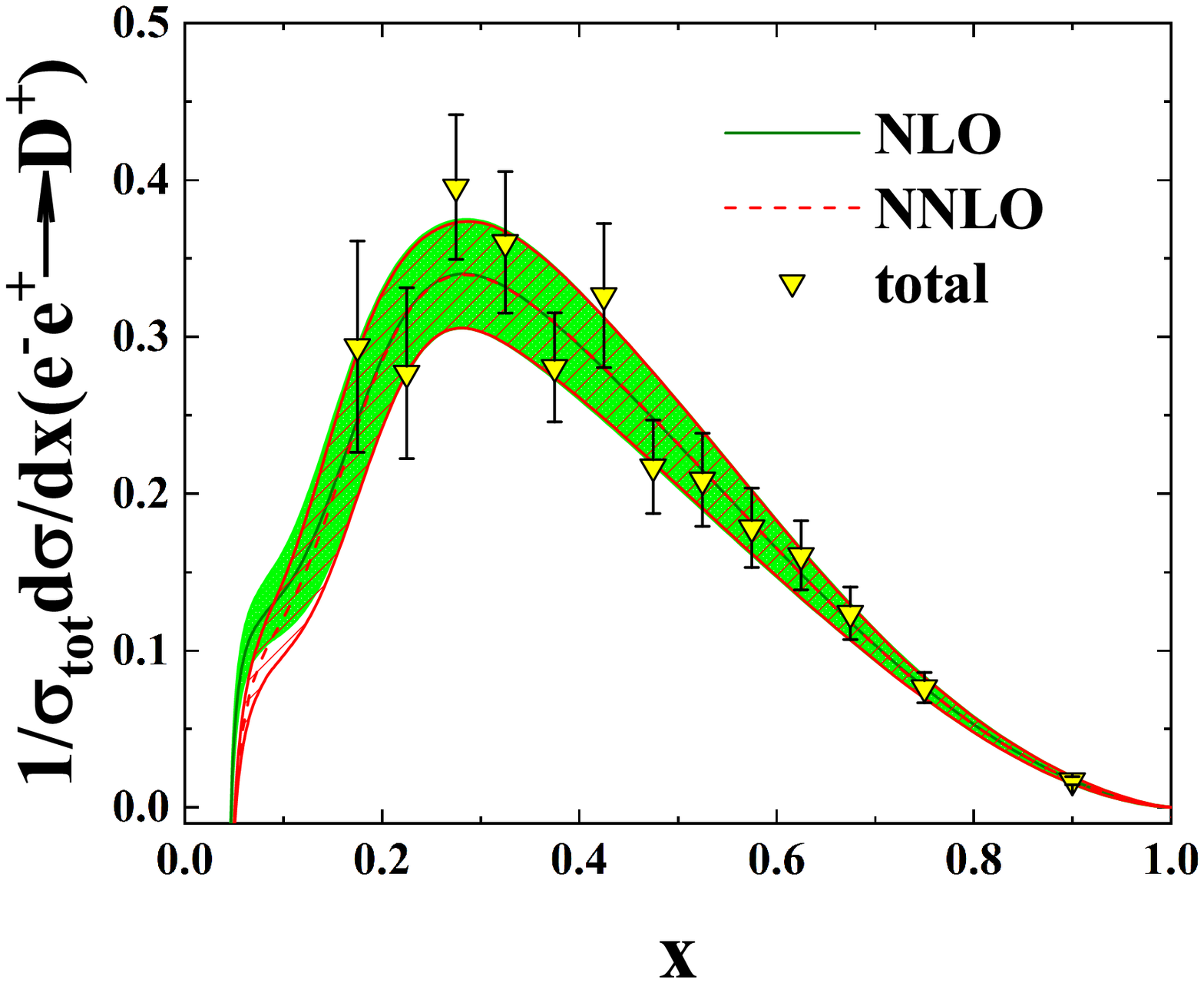}}
	\vspace{-5 cm}
	\begin{center}
		\caption{{\small Same as Fig.~\ref{fig:D0-Theory} but for the $D^+$ meson.} \label{fig:Dp-Theory}}
	\end{center}
\end{figure*}
The best values of the fit parameters for the FFs of charm and bottom quarks into the $D^0$ and $D^+$ mesons are listed in Tables.~\ref{D0pars} and \ref{D+pars} for the NLO and NNLO accuracies at the initial scale $\mu_0$. Since the time-like matching conditions are not known for NNLO correction we chose the input scale as $\mu_0=18.5$~GeV$^2$~\cite{Bertone:2017tyb,Soleymaninia:2017xhc} which is above the $b$-quark mass threshold.  
According to the Table.~\ref{tab1}, the value of $\chi^2/{\tt {d.o.f}}$ for the $D^0$ FFs at NLO and NNLO accuracies are 1.149 and 1.05, respectively. Hence, one can conclude that the inclusion of higher order QCD corrections can improve the fit quality. The same conclusion does also hold for the $D^+$ FFs analysis, see Table.~\ref{tab2}. In fact, a comparison between the values of  $\chi^2_{{\tt {NNLO}}/{\text {d.o.f}}} = 0.69$ and  $\chi^2_{{\tt {NLO}}/{\text {d.o.f}}} = 0.75$ is a reason for the goodness of fit at NNLO analysis.\\
In Fig.~\ref{fig:D0}, we plotted the FFs of  $D^0$ and $D^+$ at NLO and NNLO accuracies at higher energy scale $\mu^2=100$~GeV$^2$. To do this, we studied the $D^0$ fragmentation from  the $c+\bar{c}$ and $b+\bar{b}$ quarks as well as the gluon contribution for both NLO (solid lines) and NNLO (dashed lines) accuracies. 
For a quantitative comparison, our results are compared with the ones obtained through using the fit parameters  presented by the {\tt KKKS08} Collaboration (dot-dashed lines) \cite{Kneesch:2007ey}. 
As is seen, our result for the transition of  $b+\bar{b}$ into the $D^0$ meson is in good agreement with the corresponding result obtained from the {\tt KKKS08} analysis. 
For the FF of $c+\bar{c}$, our results are different from the  {\tt KKKS08} analysis, specifically, in the large $z$. 
For the case of gluon FF $zD_g^{D^0} (z, \mu^2)$, a  significant difference is seen for a wide range of $z$.
The error bands in Fig.~\ref{fig:D0} correspond to the choice of one-unit tolerance $\Delta \chi^2 = 1$ obtained using the Hessian approach. The green and yellow bands show the NLO and NNLO error uncertainties, respectively.
For the differences between our results and the ones from the {\tt KKKS08} analysis, it should be noted that the data sets included in our fit and the {\tt KKKS08} analysis are different. In fact, the {\tt KKKS08} Collaboration has included  the data from  {\tt BELLE} in their analyses as well, which is now removed due to an unrecoverable error in the measurements. 
Moreover, CLEO data has been also included in the {\tt KKKS08} analysis. Since, both the CLEO and Belle dataset have been measured at low energy then they cause a slightly harder FF, as is expected. According to the results presented in  tables 1 and 2 of Ref.~\cite{Kneesch:2007ey}, these two datasets impact the $c$-quark FF parameters more than the $b$-quark one.
On the other hand, the initial scales of energy are not the same in both  analyses so that the {\tt KKKS08} has set the initial scale as $\mu_0^2=m_c^2$ while ours is $\mu_0^2=18.5$~GeV$^2$ which is a little larger  than $m_b^2$. \\
In Fig.~\ref{fig:DP},  the same study is done for the $D^+$ at the scale $\mu^2 = 100$ GeV$^2$ for both NLO (solid lines) and NNLO (dashed lines) analyses. They are also  compared with the results obtained through the {\tt KKKS08} analysis (dot-dashed lines).
Our result for the $b$-quark FF is in good agreement with the one obtained by {\tt KKKS08} analysis. In the case of charm quark and gluon FFs our results behave differently in comparison  to the {\tt KKKS08} FFs.
Nevertheless, as is seen from Tables.~\ref{tab1} and \ref{tab2}, the inclusion of higher order QCD corrections leads to improve the value of $\chi^2/{\tt d.o.f}$ slightly when going from NLO to the NNLO accuracy. The central values and the size of uncertainty for the fragmentation of heavy quarks $q=c+\bar{c}, b+\bar{b}$ and gluon into the $D^+$ meson are  found similar in behavior at NLO and NNLO QCD corrections.

Concerning the gluon and light quarks FFs, remember that  there are no free fit parameters associated to their FFs so their forms were  taken to be zero at the initial scale. Available data do not provide constraints on the gluon and light quark FFs so they are generated at higher scales through the DGLAP evolution equations~\cite{DGLAP}. The small uncertainty bands for the gluon FF in Figs.~\ref{fig:D0} and \ref{fig:DP} are expected because they are only  sensitive to the quark FFs uncertainties, directly, through DGLAP evolution.
Since, the gluon FF really matters, for example, in the process $ p + p \to  D + X$ at the Tevatron or the LHC with a contribution to the differential cross section easily exceeding $50\%$, then its accurate determination as much as possible is needed. 
Hence, in order to well constrain the gluon FF, one needs to includes hadron collider observables which also have a significant impact on the gluon FF through a direct determination. However, this is beyond the scope of our present analysis and left for future work.

\subsection{ Data/theory comparisons } \label{sec:ResultsDatatheory}

In what follows we mainly focus on our NLO and NNLO theoretical  predictions for the $D^0$ and $D^+$ mesons production in $e^+ e^-$ annihilation process. Then, we compare the predictions with the inclusive total and $b$-tagged data sets to judge the goodness of our fit.
Fig.~\ref{fig:D0-Theory} includes our theoretical calculation for  the total and $b$-tagged cross sections by applying the evolved FFs at the scale $\mu =M_Z$ at NLO (solid line) and NNLO (dashed line) accuracies as well as their 1-$\sigma$ uncertainty bands.  Our  results are compared to the {\tt OPAL} data for cross section measurements at the scale $\mu=M_Z$.  As is seen, Fig.~\ref{fig:D0-Theory} shows a clear agreement between our NLO/NNLO QCD fits and the data sets which indicates a good fit quality of our QCD analyses. 
Our NLO and NNLO theoretical predictions for the normalized inclusive total and $b$-tagged  cross sections of $D^+$ meson production are also  shown in Fig.~\ref{fig:Dp-Theory}. These theoretical  predictions are also in good agreements with the analyzed data sets. With a few accuracy, it is seen that the inclusion of higher order QCD corrections up to NNLO  leads to reduction in the size of cross section at the region $x_D\le 0.2$ and for larger scaling variable $x_D$ the NNLO correction does not change the NLO theoretical predictions, significantly. The error bands of our theoretical predictions for NLO and NNLO corrections are rather the same.

\subsection{ The branching fractions and  the average energy fraction of hadrons} \label{sec:Resultsbranching fractions}

Besides the $(c, b) \rightarrow H_c$ FFs themselves, their first two moments are also of phenomenological interest. The first one corresponds to the branching fractions which is defined as
\begin{equation}\label{eq10}
B_Q(\mu)
=
\int^1_{z_{cut}}
dz D(z,\mu^2).
\end{equation}
The second moment corresponds to the average fraction of energy that the produced meson receives from the parent quark Q, namely
\begin{equation}\label{eq11}
\langle z \rangle_Q (\mu)=
\frac{1}{B_Q(\mu)} 
\int^1_{z_{cut}}
dz zD(z,\mu^2),
\end{equation}
where the cut $z_{\rm cut}=0.1$ excludes the problematic $z$ range where our formalism is not valid, however, in Figs.~\ref{fig:D0-Theory} and \ref{fig:Dp-Theory} it is shown that there are no experimental data at $z <z_{\rm cut}$. These two quantities stringently characterize   the lineshape in $x$ of the $Q\to H_c$ FFs at a given value of $\mu$ and simplify the comparisons with previous FF sets introduced by other authors.
\begin{table}[h!]
	\caption{The values of $B_c(\mu)$ and  $\langle z \rangle_c(\mu)$ for the $D^+$ and $D^0$ mesons at NLO accuracy. }
	\begin{tabular}{lccccccll}\hline\hline
		$H_c$ & & $B_c(2m_b)$ & & $B_c(m_z)$  & & $\langle z \rangle_c(2m_b)$ & & $\langle z \rangle_c(m_z)$ \\\hline
		$D^0$ & &  0.582 & & 0.561 & & 0.563 & & 0.473  \\
		$D^+$  & &  0.227 & & 0.219 & & 0.557 & & 0.468 \\ \hline \hline
	\end{tabular} \label{tab5}
\end{table}
\begin{table}[h!]
	\caption{Same as Table \ref{tab5} but at NNLO accuracy.}
	\begin{tabular}{lccccccll}\hline\hline
		$H_c$ & & $B_c(2m_b)$ & & $B_c(m_z)$  & & $\langle z \rangle_c(2m_b)$ & & $\langle z \rangle_c(m_z)$ \\\hline
		$D^0$ & &  0.593 & & 0.582 & & 0.570 & & 0.478 \\
		$D^+$  & &  0.242   & & 0.222  & & 0.570  & & 0.474 \\  \hline \hline
	\end{tabular} \label{tab6}
\end{table}
Tables.~\ref{tab5} and \ref{tab6} contain the values of $B_c$ and $\langle z \rangle_c $ at the NLO and NNLO accuracies for the $D^0$ and $D^+$ mesons, respectively, at $\mu=2m_b$ and $M_Z$. These results can be compared with the  values of the average energy fractions and the branching fractions of the $c\to D^0/D^+$ transitions measured by the {\tt OPAL} Collaboration~\cite{Alexander:1996wy,Ackerstaff:1997ki} at $\sqrt{s}=M_Z$, i.e.
\begin{eqnarray}\label{exp}
B_{c\to D^0}(M_Z)&=&0.605\pm 0.040,\nonumber\\
B_{c\to D^+}(M_Z)&=&0.235\pm 0.032,\\ 
\langle z \rangle_{c\to D^0} (M_Z)&=&0.487\pm 0.014,\nonumber\\
\langle z \rangle_{c\to D^+} (M_Z)&=&0.483\pm 0.019. \nonumber
\end{eqnarray}
A comparison between these experimental values  and the theoretical ones presented in Tables.~\ref{tab5} and \ref{tab6} shows the convenient effect of NNLO corrections which lead to  much more consistencies. We note also that at each order of perturbative QCD corrections the values of $\langle z \rangle(\mu)$ are shifted towards smaller values through the DGLAP evolution in $\mu$, as is expected.

%
\section{ Energy spectrum of charmed mesons in top quark decays}\label{sec:energy-spectrum}

As a topical application of our D-meson FFs, we study inclusive single D-meson  production at the LHC. These mesons may be produced directly or through the decay of heavier particles, including the $Z$ boson, the Higgs boson, and the top quark. For definiteness, we turn to apply the extracted $D^0/D^+$-FFs to make our phenomenological predictions for the energy distribution of charmed mesons produced in top quark decays through the following process
\begin{eqnarray}\label{pros}
t \rightarrow b + W^+ (g)
\rightarrow W^+ + D^0/D^+ + X,
\end{eqnarray}
where $X$ collectively denotes any other final-state particles. The study of energy spectrum  of produced mesons through top decays might be considered as an indirect channel to search for the top quark properties.\\
At the parton level in the process (\ref{pros}), both the $b$-quark and gluon may hadronize into the charmed mesons. The gluon fragmentation contributes to the real radiations at NLO and higher orders. Although, the contribution of gluon fragmentation can not be discriminated but it is considered to have accurate predictions for the energy spectrum of outgoing mesons. 

Ignoring the $b$-quark mass and by working in the ZM-VFN scheme, to obtain the energy distribution of charmed mesons through the process (\ref{pros})  we employ the factorization theorem  as
\begin{eqnarray}
\label{eq:master}
\frac{d\Gamma}{dx_D}
=\sum_{i=b,g}
\int_{x_i^{min}}^{x_i^{max}}
\frac{dx_i}{x_i}\,
\frac{d\Gamma}{dx_i}
(\mu_R,\mu_F)
D_i^{D^{+}}
\left(\frac{x_D}{x_i},
\mu_F\right), \nonumber \\
\end{eqnarray}
where, following Ref.~\cite{Kniehl:2012mn}, we defined the scaled-energy fraction of charmed meson as $x_D = 2 E_{D}/(m_t^{2} - m_W^{2})$. In the above relation,
$d\Gamma/dx_i $ are the differential decay rates of the process $t \to i + W^+ (i=b, g)$ at the parton level.
In Refs.~\cite{Nejad:2013fba,Kniehl:2012mn}, the NLO analytical expressions of the differential decay widths $d \Gamma/dx_i$  are presented for the polarized and unpolarized top quark decays.
In (\ref{eq:master}), the factorization ($\mu_F$) and renormalization ($\mu_R$) scales are arbitrary but to remove the large logarithms which  appear in the differential decay rate, here, we set them to $\mu_R = \mu_F = m_t = 172.9$ GeV.

\begin{figure}
	\begin{center}
		\includegraphics[width=0.85\linewidth,bb=137 42 690 690]{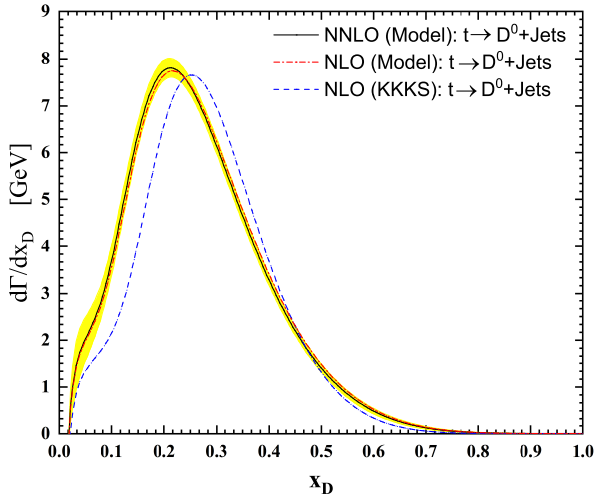}
		\caption{\label{fig1}%
			$d\Gamma(t\to D^0+Jets)/dx_D$ as a function of $x_D$ in NLO (dot-dashed line) and NNLO (solid line) accuracies at $\mu=m_t$. Our results are also compared with the NLO one from {\tt KKKS08}~\cite{Kneesch:2007ey} (dashed).}
	\end{center}
\end{figure}

\begin{figure}
	\begin{center}
		\includegraphics[width=0.85\linewidth,bb=137 42 690 690]{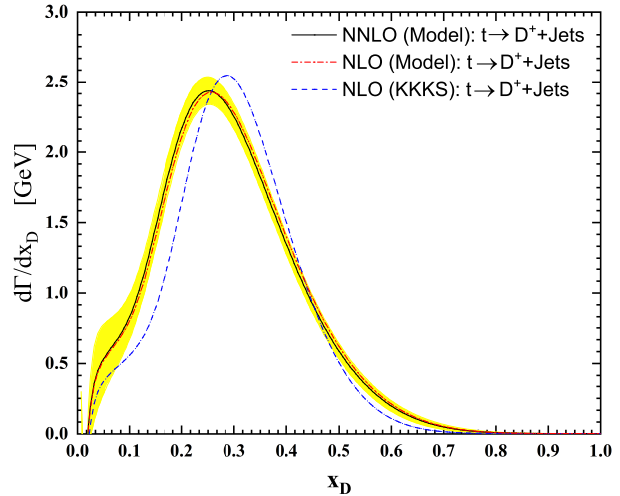}
		\caption{\label{fig2}%
			As in Fig.~\ref{fig1}, but for the $D^+$-meson.  }
	\end{center}
\end{figure}

Adopting $m_W=80.39$ GeV, in Fig.~\ref{fig1} we plotted the energy distribution of $D^0$-meson produced in the unpolarized top quark decay at NLO (dot-dashed line) and NNLO (solid line) accuracies. We have also presented an uncertainty band for the NNLO result. Using the NLO FFs of $(b,g)\to D^0$ from Ref.~\cite{Kneesch:2007ey}, we have also compared our results with the one from the {\tt KKKS08} Collaboration (dashed).
In Fig.~\ref{fig2}, we have presented our NLO and NNLO predictions for the energy distribution of $D^+$-mesons along with the uncertainty band for the NNLO result. As is seen from these figures, in comparison with the {\tt KKKS08} result our FFs lead to an enhancement(reduction) in the energy spectrum of $D^0(D^+)$ meson at the peak position. 
Study of the $x_D$-distribution (i.e. $d\Gamma/dx_D$) of the dominant decay mode $t \to W^+ +D^0/D^+ + X$ at the CERN LHC will enable us to deepen our understanding of the nonperturbative aspects of $D$-meson formation by hadronization and to pin down the $(b, g) \to D^+/D^0$ FFs. 
At the LHC, the study of this energy spectrum can be also considered as a new window towards searches on new physics. 
In Ref.~\cite{MoosaviNejad:2012ju}, Fig.~8,  the energy spectrum of B-mesons have been studied through the dominant decays of top quark in the general two Higgs doublet model (2HDM), i.e. $t\to b(\to B+X)+H^+$. There is shown that there is a considerable difference between the energy spectrum of mesons produced in the Standard Model, i.e.  $t\to b(\to B+X)+W^+$, and 2HDM. In other words, any considerable deviation of energy distribution of produced mesons from the  theoretical predictions in the SM theory can be related to the new physics, see also \cite{Abbaspour:2018ysj,MoosaviNejad:2019agw}.


%
\section{ Summary and Conclusions} \label{sec:conclusion}
%

The first set of nonperturbative FFs for bottom-flavored hadrons (B-hadrons), both at NLO and NNLO in the ZM-VFNS, was presented in our previous work~\cite{Salajegheh:2019ach} by fitting to all available experimental data of
inclusive single $B$-hadron production in $e^+e^-$ annihilation,
$e^+e^-\to B+X$, from ALEPH \cite{Heister:2001jg}, 
OPAL \cite{Abbiendi:2002vt}, SLD \cite{Abe:2002iq}, and DELPHI \cite{DELPHI:2011aa}.
In Ref.~\cite{Soleymaninia:2017xhc}, we have also provided the first QCD analysis of $(c,b)\to D^{*\pm}$ FFs up to NNLO accuracy through a QCD fit to SIA data from ALEPH~\cite{Barate:1999bg} and OPAL~\cite{Ackerstaff:1997ki} collaborations at LEP.
In both works, the effects of hadron mass were ignored.
In the present study, a new determination of nonperturbative FFs for $D^0$ and $D^+$ mesons are presented at NLO and, for the first time, at NNLO in perturbative QCD considering the hadron mass effects. Of all the available data sets for $D^0$ and $D^+$ productions in $e^+e^-$ annihilation, i.e. {\tt OPAL}, {\tt CLEO} and {\tt BELLE}, the $D^0/D^+$-FFs are determined by fitting the data taken by the {\tt OPAL} Collaboration at LEP, because the data from {\tt BELLE}  is now removed due to an unrecoverable error in the measurements and including the {\tt CLEO} data sets in the analysis might be a reason  for tension. The uncertainties of extracted FFs and SIA cross sections are also estimated using the standard `Hessian'' approach. The determination of uncertainty includes the error estimation for the tolerance criterion of $\chi^2 = 1$.  We judged the quality of our fit by comparing with the prior result presented by the {\tt KKKS08} Collaboration~\cite{Kneesch:2007ey} and also showed how they describe the available data for charmed meson productions. Considering the $\chi^2$ values, we have shown that the inclusions of higher order QCD corrections and hadron mass effects could slightly improve the fit quality.

As is seen from Figs.~\ref{fig:D0} and \ref{fig:DP}, our results for the fragmentation of  $b+\bar{b}$ into the $D^0$ and $D^+$ mesons are in good agreement with the ones presented by {\tt KKKS08}. In comparison to the {\tt KKKS08}'s results, there is a considerable difference between the {\tt KKKS08} FFs for the transition $c+\bar{c} \to D^0/D^+$ and ours at large values of $z$. This difference holds for the gluon FFs over the whole range of $z$. The reason for this inconsistency might be due to the different initial energy scales chosen, the different SIA data sets used and different schemes applied. To check the validity of our results, we have also calculated the branching fractions of mesons and the average fractions of energy carried by the outgoing mesons and compared them with the available experimental data. A good consistency between them, specifically at NNLO, guarantees the validity of our analysis.

As a typical application of extracted FFs, we employed the NLO and NNLO FFs to make our theoretical predictions for the scaled-energy distributions of $D^0/D^+$-mesons inclusively produced in top quark decays.

To describe the novelty and innovation of our work, it should be noted that this study introduces the following improvements. 
It is one of the first attempts to contain higher order QCD corrections into the $D^0/D^+$ mesons FFs considering all available data. As was explained, the results of Refs.~\cite{Kniehl:2006mw,Kneesch:2007ey}, as the only references focused on the D-mesons FFs, have included the {\tt BELLE} data in their analyses as well which are now removed from the Durham database due to some errors reported by the authors. Our work also includes the effects of hadron mass corrections into the FF analysis. Hence, the findings of this paper could make several contributions to the current literature. For future study, it would be interesting to include other sources of experimental observables which carry more information on quark and gluon FFs. 
Apart from the data, this study is possible if other theoretical partonic cross sections for single inclusive production of partons at NNLO accuracy are available. Another possible area of future research would be to investigate the electromagnetic initial-state radiation (ISR) as well as the effects of heavy quarks mass corrections, however, the latter is of minor importance. The bulk of ISR corrections is due to the effect of electromagnetic radiation emitted from the incoming electrons and positrons. As was explained in Ref.~\cite{Kneesch:2007ey}, the impact of ISR on the determination of FFs is found to be non-negligible for the analysis of the {\tt BELLE} and CLEO data which are not included in our work.

The NLO and NNLO $D^0$ and $D^+$ FFs sets presented in this work are available in the standard {\tt LHAPDF} format~\cite{Buckley:2014ana} from the author upon request.

%
\begin{acknowledgments}
%

Authors thank School of Particles and Accelerators, Institute for Research in Fundamental Sciences (IPM) for financial support of this project.
Hamzeh Khanpour also is thankful the University of Science and Technology of Mazandaran for financial support provided for this research.

\end{acknowledgments}
%




\begin{thebibliography}{}


\bibitem{Acosta:2003ax}
D.~Acosta {\it et al.} [CDF Collaboration],
``Measurement of prompt charm meson production cross sections in $p\bar{p}$ collisions at $\sqrt{s} = 1.96$ TeV,''
\href{http://dx.doi.org/10.1103/PhysRevLett.91.241804}{{\rm Phys.\ Rev.\ Lett.} {\bfseries 91}, 241804 (2003)}.



\bibitem{Derrick:1995sc}
M.~Derrick {\it et al.} [ZEUS Collaboration],
``Study of $D^{*\pm}$ (2010) production in e p collisions at HERA,''
\href{http://dx.doi.org/10.1016/0370-2693(95)00253-H}{{\rm Phys.\ Lett.\ B} {\bfseries 349}, 225 (1995)}.



\bibitem{Aid:1996hj}
S.~Aid {\it et al.} [H1 Collaboration],
``Photoproduction of $D^\pm$ mesons in electron - proton collisions at HERA,''
\href{http://dx.doi.org/10.1016/0550-3213(96)00275-1}{{\rm Nucl.\ Phys.\ B} {\bfseries 472}, 32 (1996)}.




\bibitem{CidVidal:2018eel}
X.~Cid Vidal  {\it et al.},
``Beyond the Standard Model Physics at the HL-LHC and HE-LHC,''
\href{https://arxiv.org/abs/1812.07831}{arXiv:1812.07831 [hep-ph]}.




\bibitem{Azzi:2019yne}
P.~Azzi {\it et al.} [HL-LHC Collaboration and HE-LHC Working Group],
``Standard Model Physics at the HL-LHC and HE-LHC,''
\href{https://arxiv.org/abs/1902.04070}{arXiv:1902.04070 [hep-ph]}.




\bibitem{Aktas:2004ka}
A.~Aktas {\it et al.} [H1 Collaboration],
``Inclusive production of $D^+, D^0, D^+_s$ and $D^{*+}$ mesons in deep inelastic scattering at HERA,''
\href{http://dx.doi.org/10.1140/epjc/s2004-02069-x}{{\rm Eur.\ Phys.\ J.\ C} {\bfseries 38}, 447 (2005)}.




\bibitem{Breitweg:2000qi}
J.~Breitweg {\it et al.} [ZEUS Collaboration],
``Measurement of inclusive $D^\pm_s$ photoproduction at HERA,''
\href{http://dx.doi.org/10.1016/S0370-2693(00)00431-7}{{\rm Phys.\ Lett.\ B} {\bfseries 481}, 213 (2000)}.






\bibitem{Bertone:2018ecm}
V.~Bertone {\it et al.} [NNPDF Collaboration],
``Charged hadron fragmentation functions from collider data,''
\href{http://dx.doi.org/10.1140/epjc/s10052-018-6130-4}{{\rm Eur.\ Phys.\ J.\ C} {\bfseries 78}, 651 (2018)}.


\bibitem{deFlorian:2007ekg}
D.~de Florian, R.~Sassot and M.~Stratmann,
``Global analysis of fragmentation functions for protons and charged hadrons,''
\href{http://dx.doi.org/10.1103/PhysRevD.76.074033}{{\rm Phys.\ Rev.\ D} {\bfseries 76}, 074033 (2007)}.






\bibitem{Kniehl:2000fe}
B.~A.~Kniehl, G.~Kramer and B.~Potter,
``Fragmentation functions for pions, kaons, and protons at next-to-leading order,''
\href{http://dx.doi.org/10.1016/S0550-3213(00)00303-5}{{\rm Nucl.\ Phys.\ B} {\bfseries 582}, 514 (2000)}.




\bibitem{Ethier:2017zbq}
J.~J.~Ethier, N.~Sato and W.~Melnitchouk,
``First simultaneous extraction of spin-dependent parton distributions and fragmentation functions from a global QCD analysis,''
\href{http://dx.doi.org/doi:10.1103/PhysRevLett.119.132001}{{\rm Phys.\ Rev.\ Lett.} {\bfseries 119}, 132001 (2017)}.






\bibitem{deFlorian:2007aj}
D.~de Florian, R.~Sassot and M.~Stratmann,
``Global analysis of fragmentation functions for pions and kaons and their uncertainties,''
\href{http://dx.doi.org/10.1103/PhysRevD.75.114010}{{\rm Phys.\ Rev.\ D} {\bfseries 75}, 114010 (2007)}.






\bibitem{Bertone:2017tyb}
V.~Bertone {\it et al.} [NNPDF Collaboration],
``A determination of the fragmentation functions of pions, kaons, and protons with faithful uncertainties,''
\href{http://dx.doi.org/10.1140/epjc/s10052-017-5088-y}{{\rm Eur.\ Phys.\ J.\ C} {\bfseries 77}, 516 (2017)}.







\bibitem{Kniehl:2006mw}
B.~A.~Kniehl and G.~Kramer,
``Charmed-hadron fragmentation functions from CERN LEP1 revisited,''
 \href{http://dx.doi.org/10.1103/PhysRevD.74.037502}{{\rm Phys.\ Rev.\ D} {\bfseries 74}, 037502 (2006)}.



\bibitem{Alexander:1996wy}
G.~Alexander {\it et al.} [OPAL Collaboration],
``A Study of charm hadron production in $Z^0 ---> c\bar{c}$ and $Z^0 ---> b\bar{b}$ decays at LEP,''
\href{http://dx.doi.org/10.1007/s002880050218}{{\rm Z.\ Phys.\ C} {\bfseries 72}, 1 (1996)}.





\bibitem{Ackerstaff:1997ki}
K.~Ackerstaff {\it et al.} [OPAL Collaboration],
``Measurement of $f(c ---> D^{*+} X)$, $f(b ---> D^{*+} X)$ and $\Gamma_{c\bar{c}} / \Gamma_{had}$ using $D^{*\pm}$ mesons,''
\href{http://dx.doi.org/10.1007/s100520050095}{{\rm Eur.\ Phys.\ J.\ C} {\bfseries 1}, 439 (1998)}.




\bibitem{Kneesch:2007ey}
T.~Kneesch, B.~A.~Kniehl, G.~Kramer and I.~Schienbein,
``Charmed-meson fragmentation functions with finite-mass corrections,''
\href{http://dx.doi.org/10.1016/j.nuclphysb.2008.02.015}{{\rm Nucl.\ Phys.\ B} {\bfseries 799}, 34 (2008)}.
Nucl.\ Phys.\ B {\bf 799} (2008) 34.



\bibitem{Barate:1999bg}
R.~Barate {\it et al.} [ALEPH Collaboration],
``Study of charm production in Z decays,''
\href{http://dx.doi.org/10.1007/s100520000421}{{\rm Eur.\ Phys.\ J.\ C} {\bfseries 16}, 597 (2000)}.




\bibitem{Seuster:2005tr}
R.~Seuster {\it et al.} [Belle Collaboration],
``Charm hadrons from fragmentation and B decays in $e^+ e^-$ annihilation at $\sqrt{s} = 10.6-GeV$,''
\href{http://dx.doi.org/10.1103/PhysRevD.73.032002}{{\rm Phys.\ Rev.\ D} {\bfseries 73}, 032002 (2006)}.





\bibitem{Artuso:2004pj}
M.~Artuso {\it et al.} [CLEO Collaboration],
``Charm meson spectra in $e^{+} e^{-}$ annihilation at 10.5-GeV c.m.e.,''
\href{http://dx.doi.org/10.1103/PhysRevD.70.112001}{{\rm Phys.\ Rev.\ D} {\bfseries 70}, 112001 (2004)}.







\bibitem{Kniehl:2012ti}
B.~A.~Kniehl, G.~Kramer, I.~Schienbein and H.~Spiesberger,
``Inclusive Charmed-Meson Production at the CERN LHC,''
\href{http://dx.doi.org/10.1140/epjc/s10052-012-2082-2}{{\rm Eur.\ Phys.\ J.\ C} {\bfseries 72}, 2082 (2012)}.


\bibitem{Hamon:2018zqs}
J.~Hamon [ALICE Collaboration],
``D-meson production in proton-proton collisions with ALICE at the LHC,''
\href{http://dx.doi.org/10.1016/j.nuclphysbps.2018.03.004}{{\rm Nucl.\ Part.\ Phys.\ Proc.} {\bfseries 294-296}, 32 (2018)}.



\bibitem{Adam:2016ich}
J.~Adam {\it et al.} [ALICE Collaboration],
``$D$-meson production in $p$-Pb collisions at $\sqrt{s_{\rm NN}}=$5.02 TeV and in pp collisions at $\sqrt{s}=$7 TeV,''
\href{http://dx.doi.org/10.1103/PhysRevC.94.054908}{{\rm Phys.\ Rev.\ C} {\bfseries 94}, 054908 (2016)}.


\bibitem{Maciula:2013oba}
R.~Maciula and A.~Szczurek,
``Charmed mesons and leptons from semileptonic decays at the LHC,''
\href{http://dx.doi.org/10.22323/1.191.0169}{{\rm PoS DIS} {\bfseries 2013}, 169 (2013)}.


\bibitem{Schweda:2014tya}
K.~O.~Schweda,
``Prompt production of D mesons with ALICE at the LHC,''
\href{https://arxiv.org/abs/1402.1370}{arXiv:1402.1370 [nucl-ex]}.







\bibitem{Collins:1998rz}
J.~C.~Collins,
``Hard scattering factorization with heavy quarks: A General treatment,''
\href{http://dx.doi.org/10.1103/PhysRevD.58.094002}{{\rm Phys.\ Rev.\ D} {\bfseries 58}, 094002 (1998)}.







\bibitem{Anderle:2017cgl}
D.~P.~Anderle, T.~Kaufmann, M.~Stratmann, F.~Ringer and I.~Vitev,
``Using hadron-in-jet data in a global analysis of $D^{*}$ fragmentation functions,''
\href{http://dx.doi.org/10.1103/PhysRevD.96.034028}{{\rm Phys.\ Rev.\ D} {\bfseries 96}, 034028 (2017)}.




\bibitem{Rijken:1996vr}
P.~J.~Rijken and W.~L.~van Neerven,
``$O (\alpha_s^2)$ contributions to the longitudinal fragmentation function in $e^+ e^-$ annihilation,''
\href{http://dx.doi.org/10.1016/0370-2693(96)00898-2}{{\rm Phys.\ Lett.\ B} {\bfseries 386}, 422 (1996)}.



\bibitem{Rijken:1996ns}
P.~J.~Rijken and W.~L.~van Neerven,
``Higher order QCD corrections to the transverse and longitudinal fragmentation functions in electron - positron annihilation,''
\href{http://dx.doi.org/10.1016/S0550-3213(96)00669-4}{{\rm Nucl.\ Phys.\ B} {\bfseries 487}, 233 (1997)}.
Nucl.\ Phys.\ B {\bf 487}, 233 (1997)
[hep-ph/9609377].




\bibitem{Mitov:2006wy}
A.~Mitov and S.~O.~Moch,
``QCD Corrections to Semi-Inclusive Hadron Production in Electron-Positron Annihilation at Two Loops,''
\href{http://dx.doi.org/10.1016/j.nuclphysb.2006.05.018}{{\rm Nucl.\ Phys.\ B} {\bfseries 751}, 18 (2006)}.



\bibitem{Gorishnii:1990vf}
S.~G.~Gorishnii, A.~L.~Kataev and S.~A.~Larin,
``The $O(\alpha^{3}_{s})$-corrections to $\sigma_{tot}(e^{+}e^{-}\rightarrow hadrons)$ and $\Gamma(\tau^{-} \rightarrow \nu_{\tau} + hadrons)$ in QCD,''
\href{http://dx.doi.org/10.1016/0370-2693(91)90149-K}{{\rm Phys.\ Lett.\ B} {\bfseries 259}, 144 (1991)}.










\bibitem{Bowler:1981sb}
M.~G.~Bowler,
``$e^+ e^-$ Production of Heavy Quarks in the String Model,''
\href{http://dx.doi.org/10.1007/BF01574001}{{\rm Z.\ Phys.\ C} {\bfseries 11}, 169 (1981)}.


\bibitem{DGLAP}
V.~N.~Gribov and L.~N.~Lipatov,
``Deep inelastic e p scattering in perturbation theory,''
Sov.\ J.\ Nucl.\ Phys.\  {\bf 15} (1972) 438
[Yad.\ Fiz.\  {\bf 15} (1972) 781];
G.~Altarelli and G.~Parisi,
``Asymptotic Freedom in Parton Language,''
Nucl.\ Phys.\ B {\bf 126} (1977) 298;
\href{http://dx.doi.org/10.1016/0550-3213(77)90384-4}{{\rm Nucl.\ Phys.\ B} {\bfseries 46}, 641 (1977)}.
``Calculation of the Structure Functions for Deep Inelastic Scattering and $e^+ e^-$ Annihilation by Perturbation Theory in Quantum Chromodynamics.,''
Sov.\ Phys.\ JETP {\bf 46} (1977) 641
[Zh.\ Eksp.\ Teor.\ Fiz.\  {\bf 73} (1977) 1216].




\bibitem{Tanabashi:2018oca}
M.~Tanabashi {\it et al.} [Particle Data Group],
``Review of Particle Physics,''
\href{http://dx.doi.org/10.1103/PhysRevD.98.030001}{{\rm Phys.\ Rev.\ D} {\bfseries 98}, 030001 (2018)}.




\bibitem{Bertone:2013vaa}
V.~Bertone, S.~Carrazza and J.~Rojo,
``APFEL: A PDF Evolution Library with QED corrections,''
\href{http://dx.doi.org/10.1016/j.cpc.2014.03.007}{{\rm Comput.\ Phys.\ Commun.} {\bfseries 185}, 1647 (2014)}.



\bibitem{James:1975dr}
F.~James and M.~Roos,
``Minuit: A System for Function Minimization and Analysis of the Parameter Errors and Correlations,''
\href{http://dx.doi.org/10.1016/0010-4655(75)90039-9}{{\rm Comput.\ Phys.\ Commun.} {\bfseries 10}, 343 (1975)};
F.~James,
``MINUIT Function Minimization and Error Analysis:  Reference Manual Version 94.1,''
\href{https://inspirehep.net/record/1258343/files/minuit.pdf} {CERN-D-506, CERN-D506.}


\bibitem{Soleymaninia:2018uiv}
M.~Soleymaninia, M.~Goharipour and H.~Khanpour,
``First QCD analysis of charged hadron fragmentation functions and their uncertainties at next-to-next-to-leading order,''
\href{http://dx.doi.org/10.1103/PhysRevD.98.074002}{{\rm Phys.\ Rev.\ D} {\bfseries 98}, 074002 (2018)}.




\bibitem{Soleymaninia:2019sjo}
M.~Soleymaninia, M.~Goharipour and H.~Khanpour,
``Impact of unidentified light charged hadron data on the determination of pion fragmentation functions,''
\href{http://dx.doi.org/10.1103/PhysRevD.99.034024}{{\rm Phys.\ Rev.\ D} {\bfseries 99}, 034024 (2019)}.



\bibitem{Martin:2009iq}
A.~D.~Martin, W.~J.~Stirling, R.~S.~Thorne and G.~Watt,
``Parton distributions for the LHC,''
\href{http://dx.doi.org/10.1140/epjc/s10052-009-1072-5}{{\rm Eur.\ Phys.\ J.\ C} {\bfseries 63}, 189 (2009)}.




\bibitem{Stump:2001gu}
D.~Stump, J.~Pumplin, R.~Brock, D.~Casey, J.~Huston, J.~Kalk, H.~L.~Lai and W.~K.~Tung,
``Uncertainties of predictions from parton distribution functions. 1. The Lagrange multiplier method,''
\href{http://dx.doi.org/10.1103/PhysRevD.65.014012}{{\rm Phys.\ Rev.\ D} {\bfseries 65}, 014012 (2001)}.



\bibitem{Soleymaninia:2013cxa}
M.~Soleymaninia, A.~N.~Khorramian, S.~M.~Moosavi Nejad and F.~Arbabifar,
``Determination of pion and kaon fragmentation functions including spin asymmetries data in a global analysis,''
Phys.\ Rev.\ D {\bf 88} (2013) no.5,  054019
Addendum: [Phys.\ Rev.\ D {\bf 89} (2014) no.3,  039901].


\bibitem{Nejad:2015fdh}
S.~M.~Moosavi Nejad, M.~Soleymaninia and A.~Maktoubian,
``Proton fragmentation functions considering finite-mass corrections,''
\href{http://dx.doi.org/10.1140/epja/i2016-16316-6}{{\rm Eur.\ Phys.\ J.\ A} {\bfseries 52}, 316 (2016)}.



\bibitem{Albino:2008fy}
S.~Albino, B.~A.~Kniehl and G.~Kramer,
``AKK Update: Improvements from New Theoretical Input and Experimental Data,''
\href{http://dx.doi.org/10.1016/j.nuclphysb.2008.05.017}{{\rm Nucl.\ Phys.\ B} {\bfseries 803}, 42 (2008)}.



\bibitem{Albino:2005me}
S.~Albino, B.~A.~Kniehl and G.~Kramer,
``Fragmentation functions for light charged hadrons with complete quark flavor separation,''
\href{http://dx.doi.org/10.1016/j.nuclphysb.2005.07.010}{{\rm Nucl.\ Phys.\ B} {\bfseries 725}, 181 (2005)}.

\bibitem{Soleymaninia:2017xhc}
M.~Soleymaninia, H.~Khanpour and S.~M.~Moosavi Nejad,
``First determination of $D^{*+}$-meson fragmentation functions and their uncertainties at next-to-next-to-leading order,''
\href{http://dx.doi.org/10.1103/PhysRevD.97.074014}{{\rm Phys.\ Rev.\ D} {\bfseries 97}, 074014 (2018)}.



\bibitem{Kniehl:2012mn}
B.~A.~Kniehl, G.~Kramer and S.~M.~Moosavi Nejad,
``Bottom-Flavored Hadrons from Top-Quark Decay at Next-to-Leading order in the General-Mass Variable-Flavor-Number Scheme,''
\href{http://dx.doi.org/10.1016/j.nuclphysb.2012.05.008}{{\rm Nucl.\ Phys.\ B} {\bfseries 862}, 720 (2012)}.


\bibitem{Nejad:2013fba}
S.~M.~Moosavi Nejad,
``Energy spectrum of bottom- and charmed-flavored mesons from polarized top quark decay $t(\uparrow)\rightarrow W^++B/D+X$ at $O(\alpha_s)$,''
\href{http://dx.doi.org/10.1103/PhysRevD.88.094011}{{\rm Phys.\ Rev.\ D} {\bfseries 88}, 094011 (2013)}.

\bibitem{MoosaviNejad:2012ju}
S.~M.~Moosavi Nejad,
``${\cal O}(\alpha_s)$ corrections to the B-hadron energy distribution of the top decay in the Minimal Supersymmetric Standard Model considering GM-VFN scheme,''
Eur.\ Phys.\ J.\ C {\bf 72} (2012) 2224.



\bibitem{Abbaspour:2018ysj}
S.~Abbaspour, S.~M.~Moosavi Nejad and M.~Balali,
``Indirect search for light charged Higgs bosons through the dominant semileptonic decays of top quark $t\to b(\to B/D+X)+H^+(\to \tau^+\nu_\tau)$,''
Nucl.\ Phys.\ B {\bf 932} (2018) 505.

\bibitem{MoosaviNejad:2019agw}
S.~M.~Moosavi Nejad, S.~Abbaspour and R.~Farashahian,
``Interference effects for the top quark decays $t\to b+W^+/H^+(\to\tau^+\nu_\tau)$,''
Phys.\ Rev.\ D {\bf 99} (2019) no.9,  095012.

\bibitem{Salajegheh:2019ach}
M.~Salajegheh, S.~M.~Moosavi Nejad, H.~Khanpour, B.~A.~Kniehl and M.~Soleymaninia,
``$B$-hadron fragmentation functions at next-to-next-to-leading order from a global analysis of $e^+e^-$ annihilation data,''
Phys.\ Rev.\ D {\bf 99} (2019) no.11,  114001.


\bibitem{Heister:2001jg} 
A.~Heister {\it et al.} [ALEPH Collaboration],
Phys.\ Lett.\ B {\bf 512}, 30 (2001).

\bibitem{Abbiendi:2002vt} 
G.~Abbiendi {\it et al.} [OPAL Collaboration],
Eur.\ Phys.\ J.\ C {\bf 29}, 463 (2003).

\bibitem{Abe:2002iq} 
K.~Abe {\it et al.} [SLD Collaboration],
Phys.\ Rev.\ D {\bf 65}, 092006 (2002).
Erratum: [Phys.\ Rev.\ D {\bf 66}, 079905 (2002)]

\bibitem{DELPHI:2011aa} 
J.~Abdallah {\it et al.} [DELPHI Collaboration],
Eur.\ Phys.\ J.\ C {\bf 71}, 1557 (2011).



\bibitem{Buckley:2014ana}
A.~Buckley, J.~Ferrando, S.~Lloyd, K.~Nordström, B.~Page, M.~Rüfenacht, M.~Schönherr and G.~Watt,
``LHAPDF6: parton density access in the LHC precision era,''
\href{http://dx.doi.org/10.1140/epjc/s10052-015-3318-8}{{\rm Eur.\ Phys.\ J.\ C} {\bfseries 75}, 132 (2015)}.




\end{thebibliography}
\end{document}